\documentclass[aps,prd,preprint,floats,floatfix,showpacs,showkeys,superscriptaddress,nofootinbib,tightenlines]{revtex4-1}

\usepackage[dvips]{graphicx,color}
\usepackage{bm}
\usepackage{dcolumn}
\usepackage{epsfig}
\usepackage{natbib}

\newcommand{\ket}[1]{|{#1}\rangle}
\newcommand{\bra}[1]{\langle{#1}|}
\newcommand{\inp}[2]{\langle{#1}|{#2}\rangle}

\begin{document}
\bibliographystyle{apsrev4-1}

\title{Unitary coupled-channels model for three-mesons decays of heavy mesons}

\author{H. Kamano}
\affiliation{Research Center for Nuclear Physics, Osaka University, Ibaraki, Osaka 567-0047, Japan}
\affiliation{Excited Baryon Analysis Center (EBAC), Thomas Jefferson National Accelerator Facility, Newport News, Virginia 23606, USA}
\author{S. X. Nakamura}
\affiliation{Excited Baryon Analysis Center (EBAC), Thomas Jefferson National Accelerator Facility, Newport News, Virginia 23606, USA}
\author{T.-S. H. Lee}
\affiliation{Physics Division, Argonne National Laboratory, Argonne, Illinois 60439, USA}
\affiliation{Excited Baryon Analysis Center (EBAC), Thomas Jefferson National Accelerator Facility, Newport News, Virginia 23606, USA}
\author{T. Sato}
\affiliation{Department of Physics, Osaka University, Toyonaka, Osaka 560-0043, Japan}
\affiliation{Excited Baryon Analysis Center (EBAC), Thomas Jefferson National Accelerator Facility, Newport News, Virginia 23606, USA}

\date{\today}

\begin{abstract}
A unitary coupled-channels model is presented for investigating the decays of heavy mesons
and excited meson states into three light pseudoscalar mesons.
The model accounts for the three-mesons final state interactions in the decay processes,
as required by both the three-body and two-body unitarity conditions.
In the absence of the $Z$-diagram mechanisms that are necessary consequences of the three-body unitarity, 
our decay amplitudes are reduced to a form similar to those used in the so-called isobar-model analysis.
We apply our coupled-channels model to the three-pions decays of $a_1(1260)$, $\pi_2(1670)$,
$\pi_2(2100)$, and $D^0$ mesons, and show that the $Z$-diagram mechanisms can
contribute to the calculated Dalitz plot distributions by as much as  30\% in magnitudes in the regions
where $f_0(600)$, $\rho(770)$, and $f_2(1270)$ dominate the distributions.
Also, by fitting to the same Dalitz plot distributions, we demonstrate that the decay amplitudes 
obtained with the unitary model and the isobar model can be
rather different, particularly in the phase that plays a crucial role
in extracting the Cabibbo-Kobayashi-Maskawa $CP$-violating phase from the data of $B$ meson decays.
Our results indicate that the commonly used isobar-model analysis
must be extended to account for the final state interactions required by
the three-body unitarity to reanalyze the three-mesons decays of heavy mesons, 
thereby exploring hybrid or exotic mesons, and signatures of physics beyond the standard model.
\end{abstract}

%13.25.-k 	Hadronic decays of mesons
%14.40.Rt 	Exotic mesons
%11.80.Jy 	Many-body scattering and Faddeev equation
%11.80.Et 	Partial-wave analysis
%11.80.Gw 	Multichannel scattering
%11.80.La 	Multiple scattering


\maketitle

\section{\label{sec:int}Introduction}

\begin{figure}[t]
\includegraphics[width=\textwidth]{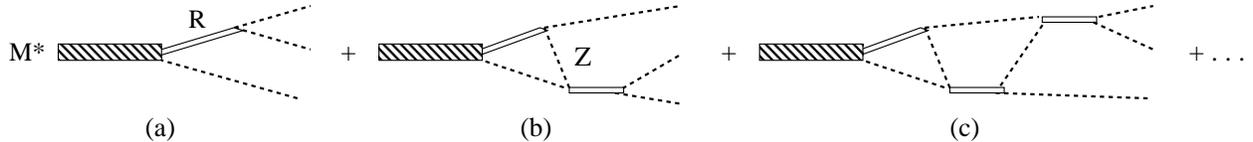}
\caption{\label{fig:mstar-decay3}
$M^*$-decay amplitude.}
\end{figure}

It has long been recognized that hadrons lying outside of the conventional 
constituent quark model must exist within the framework of Quantum Chromodynamics (QCD).
These so-called ``exotic'' hadrons,  speculated as tetra-quark states or
hybrid states or glueballs, have been predicted by various calculations 
using the Lattice QCD, the QCD sum rule, and the flux-tube model, as reviewed in Ref.~\cite{klempt}.
Thus, quite a few experimental programs have been developed to search for exotic mesons 
via the three-mesons production reactions, such as
$\pi N \to M^* N\to \pi\pi\pi N$~\cite{e852,chung,dzierba}, 
$\gamma N \to M^*N \to \pi\pi\pi N, \pi K\bar{K} N$~\cite{gluex,nozar},
and $N \bar{N} \to M^* \to \pi\pi\eta$~\cite{lear},
where the intermediate mesons $M^*$ could be exotic.
To identify $M^*$, the main task is to extract the partial-wave
amplitudes from the final three-mesons distributions.
So far, this has been done mainly by using the isobar model, within which 
two of the three mesons form a light flavor excited meson $R$ ($f_0, \rho, K^*$, etc.)
and the third meson is treated as a spectator in the decays of $R$, 
as illustrated in Fig.~\ref{fig:mstar-decay3}(a). 
There, the propagation of  $R$ is commonly described with the Breit-Wigner 
parametrization or with the two-body unitary $K$-matrix parameterizations~\cite{au,guo}
constrained by the dispersion relations.
In any case, the three-body unitarity is missing in those analyses.
The noninteracting $cR$ amplitudes, where $c$ is a spectator light pseudoscalar meson, 
and an appropriately parametrized nonresonant 
amplitude are then summed coherently with multiplicative complex parameters which are
adjusted to fit the Dalitz plot of the measured three-mesons distributions.

In an isobar-model analysis of
the $\pi^-p \to \pi^-\pi^-\pi^+ p$ and $\pi^- p\to ~\pi^-\pi^0\pi^0 p$ data
from the E852 experiment~\cite{e852,chung,dzierba}, 
$a_1(1260)$, $a_2(1320)$, $\pi_2(1670)$ and $a_4(2020)$ resonances
were identified, and the exotic $J^{PC}=1^{-+}$ meson near $1.6$ GeV [$\pi_1(1600)$] 
proposed from an earlier analysis was ruled out. 
The CLAS analysis~\cite{nozar} of $\gamma p \to \pi^+\pi^+\pi^- n$
data identified $a_2(1320)$ and $\pi_2(1670)$, but neither $a_1(1260)$ nor exotic
$\pi_1(1600)$ at the expected levels.
On the other hand, the COMPASS experiment~\cite{compass} claimed to have 
observed $\pi_1(1600)$ in the $\pi^-\pi^-\pi^+$ final state from 
a pion scattering on a lead target.
As a step to understand  the differences between the results from
these analyses as well as from the previous isobar-model analyses, 
it is necessary to first examine the extent to which the isobar model is valid. 
This is also needed for developing a theoretically sound approach to analyze the 
three-mesons photoproduction data that will be obtained
at JLab with the 12-GeV upgrade~\cite{gluex}.

The isobar model has also been commonly used to analyze the data of three-mesons decays of 
$J/\psi$~\cite{bes,babar-j}, 
$D$~\cite{d-a1,d-a2,d-a3,d-a4,cleo,babar-d0,babar-d1,babar-b6,babar-b7,belle-b6,belle-b7} and 
$B$~\cite{belle-b1,belle-b2,belle-b31,belle-b32,belle-b4,belle-b5,
cleo-b1,babar-b1,babar-b2,babar-b3,babar-b4,babar-b5} mesons.
The $B$ and $D$ decays have been analyzed with
interests in the $CP$ violation and physics beyond the standard model.
Some $B$ decay processes have also been analyzed using dispersion relations~\cite{el-bennich}, 
neglecting the interactions between the outgoing two-mesons
subsystem and the third meson, as assumed in the isobar model.
The strong phases arising from the final state interactions in the decay processes
are essential in determining the weak decay amplitudes of these heavy mesons and searching
for physics beyond the standard model.
For example, BABAR~\cite{babar-d0,babar-b6,babar-b7} and Belle~\cite{belle-b6,belle-b7} extracted the 
Cabibbo-Kobayashi-Maskawa (CKM) $CP$ violating phase  $\gamma$ from the data of
$B^{\mp} \to D^0 $ (or $\bar{D}^0$) $K^\mp \to (K_S^0 \pi^+ \pi^-) K^{\mp}$. 
They utilized the fact that the interference between the decay amplitude of
$B^{\mp} \to D^0 K^{\mp} \to (K_S^0 \pi^+ \pi^-) K^{\mp}$ and that
of $B^{\mp} \to \bar{D}^0 K^- \to (K_S^0 \pi^+ \pi^-) K^{\mp}$ is proportional
to $e^{\mp i\gamma}$.  Clearly, the accuracy of the phases of the  partial-wave
amplitudes of $ D^0 (\bar{D}^0) \to (K_S^0 \pi^+ \pi^-)$, which were determined 
within the isobar model, is crucial in extracting this fundamental parameter
$\gamma$ from the data. 
In the isobar model, the strong phases from the final state interactions
are partly accounted for by using complex $D\to \pi R$ couplings.
However, the phases of the amplitudes generally depend on
kinematics and have to satisfy the three-body unitarity, 
which is beyond what the isobar model can achieve.

The above discussions strongly indicate the need for investigating the extent 
to which the isobar model is valid. 
Within the well-developed three-hadron scattering models, as reviewed in 
Refs.~\cite{thomas,garcilazo,msl}, the isobar model is clearly a simplification 
since one of the mesons from the decay of the propagating resonance $R$ can
interact with the third meson to form another $R$. This interaction is
traditionally called the $Z$-diagram, as illustrated in Fig.~\ref{fig:mstar-decay3}(b).
It was well established in the studies of $\pi N$~\cite{thomas}, $\pi NN$~\cite{garcilazo}
and $\pi\pi N$~\cite{msl,ebac,juelich} systems that the multiple scattering due to
the $Z$-diagram mechanisms, as illustrated in Fig.~\ref{fig:mstar-decay3}(c), 
is essential to preserve the three-body unitarity for interpreting the data correctly.
Only very limited similar attempts have been 
taken recently to analyze the three-mesons decay of heavy mesons~\cite{guimaraes2,guimaraes3}.

The main purpose of this work is to apply the unitary approach developed in Ref.~\cite{msl} 
(hereafter referred to as MSL) 
to investigate the importance of the $Z$-diagram mechanisms in analyzing the data of three-mesons 
decays from heavy mesons and excited meson states.
We will present a model that satisfies the two-body and three-body unitarity conditions. 
We start with a model Hamiltonian defined by (bare) vertex interactions
$f_{ab,R}$ and $\Gamma_{cR,M^*}$ and two-body interactions $v_{c'R',cR}$, where 
$a,b,c$ are physical light pseudoscalar mesons ($\pi$, $K$ etc.),
$R$ is a light flavor excited state decaying to two light pseudoscalar mesons, $R=f_0, \rho, f_2,K^\ast,...$, 
and $M^\ast$ is a heavy meson decaying to three light pseudoscalar mesons.
The vertex interactions $f_{ab,R}$ are determined by fitting the empirical 
$ab\to ab$ scattering amplitudes, and are used to define 
the propagation of $R$ and to calculate the one-particle-exchange $Z$-diagram amplitudes $Z_{c'R',cR}$. 
The $cR \to c'R'$ scattering amplitudes $T_{c'R',cR}$ are then calculated from 
$Z_{c'R',cR}$ by solving a set of coupled-channels equations 
to account for the three-mesons final state interactions of heavy-meson decays.
In the absence of the $Z$-diagram mechanisms, our decay amplitudes are reduced to
a form similar to those used in the isobar-model analysis.
Thus we will be able to examine the effects of $Z$-diagram mechanisms
in determining the Dalitz plots and the parameters of resonances
that decay strongly into three-mesons.
The model is applied to investigate the three-pions decays of 
$a_1(1260)$, $\pi_2(1670)$, $\pi_2(2100)$, and $D^0$ mesons.

The organization of this paper is as follows.
In Sec.~\ref{sec:frm}, we present our model Hamiltonian and 
describe the derivation of a set of coupled-channel equations for
calculating the meson-$R$ scattering amplitudes from the
$Z$-diagram mechanisms, and how these amplitudes are
used to calculate the three-mesons final state interactions in heavy-meson decays.
The procedures for applying our model in practical calculations are given in Sec.~\ref{sec:nmr}.
The results for calculations of the decays of $a_1(1260)$, $\pi_2(1670)$,
$\pi_2(2100)$, and $D^0$ mesons are presented in Sec.~\ref{sec:appl}.
Summary and outlook are given in Sec.~\ref{sec:summary}.

\section{\label{sec:frm}Formulation}

Following the MSL formulation~\cite{msl} of  hadron reactions, we assume that the decays 
of heavy mesons into three mesons can be described by the following Hamiltonian,
\begin{equation}
H= H_0 + H',
\label{eq:h-tot}
\end{equation}
where $H_0$ is the free Hamiltonian of the considered degrees of
freedom: the bare heavy mesons $M^* = a_1 , \pi_2, D^0,...$, 
the bare light flavor excited mesons $R = f_0,\rho,f_2,...$,
and the physical ground pseudoscalar mesons denoted as 
$a, b, c =$ $\pi $, $K$ etc. 
The interaction Hamiltonian $H'$ is defined as (In this section, the summation
runs over the momentum, spin, and isospin spaces of the particles.),
\begin{eqnarray}
H' &=& \sum_{M^\ast}\sum_{cR}[\Gamma_{cR,M^*} + \Gamma^\dagger_{cR,M^*}] + H'' ,
\label{eq:h-int}
\end{eqnarray}
\begin{eqnarray}
H'' &=&
\sum_{c'R',cR} v_{c'R',cR} + \sum_{R}\sum_{ab}[f_{ab,R}+f^\dagger_{ab,R}] , 
\label{eq:h-int2}
\end{eqnarray}
where $v_{c'R',cR}$ denotes the $cR\to c'R'$ transition potentials;
$\Gamma_{cR,M^\ast}$ ($f_{ab,R}$) is the bare vertex describing 
$M^\ast\to cR$ ($R\to ab$) processes.
Here we note that the term $H''$ does not include any interactions with
the $M^*$ states and we have neglected the interactions between the particles ($a$,$b$) in the decay
channels of the light flavor excited meson states $R$.
Throughout this paper, we will use the ``right-to-left'' ordering for the channel indices.
(Note that $\Gamma_{M^\ast,cR}= \Gamma^\dag_{cR,M^\ast}$ and $f_{R,ab}= f^\dag_{ab,R}$ for the bare vertices.)

Starting with Eq.~(\ref{eq:h-tot}), the reaction $T$-matrix is defined by the following  equation, 
\begin{equation}
T(E)=  H' + H' \frac{1}{E-H + i\epsilon} H',
\label{eq:low}
\end{equation}
where $E$ is the total scattering energy in the center-of-mass system.
Since the considered  Hamiltonian is hermitian and energy independent,
it is straightforward to show that the $S$-matrix
$S(E)= 1 - 2\pi i \delta(E-H_0) T(E)$ is unitary 
$S^\dagger(E)S(E)=1$. This is the simplicity of this formulation to
have a unitary reaction model. 
To solve Eq.~(\ref{eq:low}), it is convenient to first define a scattering equation for
calculating the effects only from the non-$M^*$ Hamiltonian $H^{''}$ on
the scattering of the ground pseudoscalar mesons $c=\pi, K$ from
the light flavor excited meson states $R= f_0, \rho, f_2,...$.
Namely, we will first calculate the amplitude 
\[
T'_{c'R',cR}(E) =\bra{c'R'}T'(E)\ket{cR}
\]
where the non-$M^*$ scattering operator $T'(E)$  is defined by
\begin{equation}
T'(E)=  H'' + H''\frac{P}{E-\bar H + i\epsilon} H''  ,
\label{eq:low'}
\end{equation}
with $\bar H \equiv H_0 + H''$.
The intermediate states in the above equation 
are restricted by the projection operator $P$  defined by
\begin{eqnarray}
P = \sum_{cR}\ket{cR}\bra{cR} +\sum_{abc} \ket{abc}\bra{abc} .
\end{eqnarray}
By further applying the standard projection operator method~\cite{feshbach1,feshbach2},
as detailed in  Ref.~\cite{msl} for a $\pi\pi N$ Hamiltonian, one can
cast Eq.~(\ref{eq:low'}) into a form for practical calculations
of $T'_{c'R',cR}(E)$.
By simply changing the particle labels and dropping the contributions from 
the direct $\gamma_{13}$, $v_{23}$, $v_{33}$
interactions in the Appendix~B of Ref.~\cite{msl},
we can obtain the scattering amplitudes for this investigation.
The resulting $cR \to c'R'$ amplitudes, which describe the
multiple scattering mechanisms followed by a $M^\ast$ decay as illustrated
in Figs.~\ref{fig:mstar-decay3}(b) and~\ref{fig:mstar-decay3}(c), are defined by
\begin{equation}
T'_{c'R',cR}(E) = V_{c'R',cR}(E) +
\sum_{c'''R''',c''R''}  V_{c'R',c'''R'''}(E) G_{c'''R''',c''R''}(E) T'_{c''R'',cR}(E) .
\label{eq:rpi}
\end{equation}
Here the driving term is
\begin{equation}
V_{c'R',cR}(E) = v_{c'R',cR} + Z_{c'R',cR}(E),
\label{eq:rpi-v}
\end{equation}
where $v_{c'R',cR}$ is the $cR \to c'R'$ transition potential, 
and the second term is the $Z$-diagram defined with the $R\to ab$ vertex as
\begin{equation}
Z_{c'R',cR}(E) = \sum_{c''} 
f_{R', cc''}  \frac{1}{E - E_c - E_{c'} - E_{c''} +i\epsilon}  f_{c'c'',R} .
\label{eq:z}
\end{equation}
Here $c''$ is the exchanged meson.
We have also introduced a notation $E_c=\sqrt{m_c^2 + \vec{p}^2_c}$ 
to denote the free energy operator for a particle $c$ with mass $m_c$ and momentum  $\vec{p}_c$.

The Green function in Eq.~(\ref{eq:rpi}) is defined by
\begin{equation}
[G^{-1}(E)]_{c'R',cR} = 
\delta_{c',c}\left[ (E - E_c - E_R)\delta_{R',R} - \Sigma_{R',R}(E - E_c) \right] .
\label{eq:pir-G0}
\end{equation}
The self-energy of the propagation of $R$ in Eq.~(\ref{eq:pir-G0}) is
\begin{equation}
\Sigma_{R',R}(w) = 
\sum_{ab}
\bra{R'} f_{R',ab} \frac{{\cal B}_{ab}}{w- E_{a}- E_{b}+i\epsilon} f_{ab,R}\ket{R} ,
\label{eq:r-selfe}
\end{equation}
where ${\cal B}_{ab}$ is a factor associated with the Bose symmetry of mesons: 
${\cal B}_{ab} = 1/2$ if $a$ and $b$ are the identical particles or otherwise ${\cal B}_{ab}= 1$.

The self-energy~(\ref{eq:r-selfe}) also determines the $ab\to a'b'$ scattering amplitudes.
In the center-of-mass system, it has the familiar form
\begin{equation}
T_{a'b',ab}(w) = \sum_{R,R'} 
({\cal B}_{a'b'})^{1/2}f_{a'b',R'}  [d^{-1}(w)]_{R',R} ({\cal B}_{ab})^{1/2}f_{R,ab},
\label{eq:pipit}
\end{equation}
with
\begin{equation}
[d(w)]_{R',R} = (w-m_R) \delta_{R',R} - \Sigma_{R',R}(w) .
\label{eq:r-selfe2}
\end{equation}
We thus can determine the mass $m_R$ of bare $R$ state and the vertex interaction ${f}_{ab,R}$
by fitting the empirical amplitudes of the meson-meson scatterings such as 
$\pi\pi\to \pi\pi$ and $\pi K\to\pi K$. 
This then allows us to predict  the $Z$-diagram effects
on $T'_{c'R',cR}$ through solving Eq.~(\ref{eq:rpi}).

The transition potential $v_{c'R',cR}$ can be derived from phenomenological Lagrangian
by using the method of unitary transformation~\cite{ut-sato,msl}.
It can also be taken from more fundamental modelings within QCD.
This is beyond the scope of this paper, and we set $v_{c'R',cR}=0$ in solving Eq.~(\ref{eq:rpi}).
Thus the final three-mesons scattering effects predicted in this work are
only the necessary consequence of meson-meson scattering under the three-body unitarity condition.

The amplitude for the three-mesons decay of $M^\ast$,  $M^* \to abc$, is
\begin{eqnarray}
T_{abc, M^*}(E) = \bra{\Psi^{(-)}_{abc}(E)} H' \ket{M^*} ,
\label{eq:decay-t}
\end{eqnarray}
where the three-mesons scattering wave function is defined by
\begin{eqnarray}
\bra{\Psi^{(-)}_{abc}(E)} = \bra{abc}\left[1 + H''\frac{1}{E-\bar H  + i\epsilon} \right] ,
\label{eq:3pi-wf0}
\end{eqnarray}
with $\bra{abc}$ being the three-mesons plane-wave state.
From Eqs.~(\ref{eq:h-int}) and~(\ref{eq:h-int2}), we see that
\begin{eqnarray*}
H' \ket{M^*}  =  \sum_{cR}\ket{cR}\bra{cR}\Gamma_{cR,M^\ast}\ket{M^*} ,
\end{eqnarray*}
\begin{eqnarray*}
\bra{abc}H''  =  \sum^{\text{cyclic}}_{(a'b'c')}\sum_{R}\bra{abc}f_{a'b',R}\ket{c'R}\bra{c'R} . 
\end{eqnarray*}
Here the symbol $\displaystyle \sum^{\text{cyclic}}_{(a'b'c')}$ means taking summation over 
the cyclic permutation, $(a'b'c') = (abc) , (cab) , (bca) $.
Because of the orthogonality conditions, $\inp{cR}{M^*}=0$ and $\inp{abc}{cR}=0$, 
the above relations allow us to write Eq.~(\ref{eq:decay-t}) as
\begin{eqnarray}
T_{abc, M^*}(E) =\bra{abc} H'' \left[ P_{D} \frac{1}{E-\bar H+i\epsilon}P_D \right] H' \ket{M^*} ,
\label{eq:decay-tf}
\end{eqnarray}
where $P_D$ is the projection operator for the space spanned by $cR$ states,
\begin{eqnarray}
P_{D} = \sum_{cR} \ket{cR}\bra{cR} .
\label{eq:cr-proj}
\end{eqnarray}
Following the procedures in the Appendix~B of Ref.~\cite{msl}, one can show that
\begin{eqnarray}
P_{D} \frac{1}{E-\bar H+i\epsilon}P_{D}
&=& 
\sum_{c'R',cR} \ket{c'R'} G_{c'R',cR}(E) \bra{cR}
\nonumber \\
&& 
+
\sum_{cR,c'R'} \sum_{c'''R''',c''R''} 
\ket{c'R'} G_{c'R',c'''R'''}(E) T'_{c'''R''',c''R''}(E) G_{c''R'',cR}(E) \bra{cR} ,
\nonumber\\
\label{eq:3pi-wf1}
\end{eqnarray}
where $T'_{c'''R''',c''R''}(E)$ and $G_{cR,c'R'}(E)$ 
have been defined in Eqs.~(\ref{eq:rpi}) and~(\ref{eq:pir-G0}), respectively.

Substituting Eq.~(\ref{eq:3pi-wf1}) into Eq.~(\ref{eq:decay-tf})
and using the vertex functions of $H'$ defined by Eq.~(\ref{eq:h-int}),
we can write 
\begin{eqnarray}
T_{abc, M^*}(E) = \sum^{\text{cyclic}}_{(a'b'c')} T_{(a'b')c',M^\ast}(E),
\label{eq:decay-tf0}
\end{eqnarray}
with $T_{(ab)c, M^*}(E)$ being the amplitude for the subsequent decay of $M^\ast \to R c \to (ab)c$
expressed as
\begin{eqnarray}
T_{(ab)c,M^*}(E) = T^{\text{Isobar}}_{(ab)c,M^*}(E) + T^{\text{FSI}}_{(ab)c,M^*}(E) ,
\label{eq:decay-amp0}
\end{eqnarray}
where FSI stands for final state interaction, and
\begin{eqnarray}
T^{\text{Isobar}}_{(ab)c,M^*}(E) &=&
\sum_{R} \sum_{c'R'} \bra{ab}f_{ab,R} G_{cR,c'R'}(E) \Gamma_{c'R',M^*}\ket{M^*} , 
\label{eq:t-isobar}
\end{eqnarray}
\begin{eqnarray}
T^{\text{FSI}}_{(ab)c,M^*}(E) &=&
\sum_{R}\sum_{c'R'} \sum_{c'''R''',c''R''} \bra{ab} f_{ab,R} G_{c R,c'R'}(E) 
T'_{c'R', c'''R'''}(E) G_{c'''R''',c''R''}(E) \Gamma_{c''R'',M^*} \ket{M^*} .
\nonumber\\
\label{eq:decay-int}
\end{eqnarray}

\begin{figure}[t]
 \includegraphics[width=0.66\textwidth]{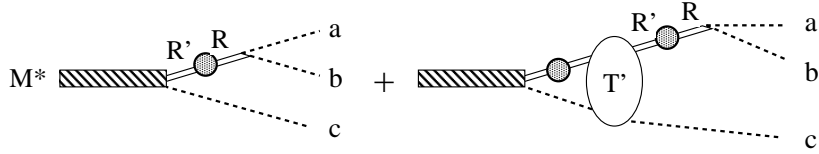}
\caption{\label{fig:mstar-decay}
Amplitude for the three-mesons decays of $M^*$ 
in the unitary coupled-channels model. 
The bulb labeled $T'$ is the $T$-matrix element
for the $cR \to c'R'$ process without $M^\ast$ excitation  [Eq.~(\ref{eq:rpi})].
The dressed $cR$ Green function, yielding $G_{cR,cR'}(E)$,
is indicated by the gray circle [Eq.~(\ref{eq:pir-G0})].
}
\label{fig:decay-t}
\end{figure}

Equation~(\ref{eq:decay-amp0}) is illustrated in Fig.~\ref{fig:decay-t}.
We now note that the commonly used isobar-model analysis corresponds to keeping 
only the term $T^{\text{Isobar}}_{abc,M^*}(E)$ within our formulation. 
The difference between different isobar-model analyses is in the parametrization
of the Green function $G_{c R,c'R'}(E)$ and the vertex function $f_{ab,R}$. 
We also note that even in this simplified case, $G_{cR,c'R'}(E)$  and $f_{ab,R}$ 
are related through Eqs.~(\ref{eq:pir-G0}) and~(\ref{eq:r-selfe})
within our formulation, but are often not treated consistently in the isobar-model analysis.

The full decay amplitude~(\ref{eq:decay-amp0}) can be concisely written as
\begin{eqnarray}
T_{(ab)c,M^*}(E) = 
\sum_{R} \sum_{c'R'} \bra{ab} f_{ab,R} G_{cR,c'R'}(E)\ket{\bar{\Gamma}_{c'R',M^*}} ,
\label{eq:decay-ampf}
\end{eqnarray}
where the dressed $M^*\to c R$ vertex function is defined by
\begin{eqnarray}
\ket{\bar{\Gamma}_{cR,M^*}}
= \sum_{c''R''} \left[\delta_{cR,c''R''} + \sum_{c'''R'''}T'_{cR,c'''R'''} G_{c'''R''',c''R''}(E)\right]
\Gamma_{c''R'',M^*}\ket{M^*}.
\label{eq:mf-dressed}
\end{eqnarray}
Obviously, $T^{\text{Isobar}}_{(ab)c,M^*}(E)$ of Eq.~(\ref{eq:t-isobar}) can be 
obtained from Eq.~(\ref{eq:decay-ampf}) by 
replacing $\ket{\bar{\Gamma}_{c'R',M^*}}$ with $\Gamma_{c'R',M^*}\ket{M^*}$.

For strong decays, the resonance pole positions and decay widths of heavy mesons 
($M^*$) can be shifted by three-mesons scattering. This can be seen by considering
the $M^*$ propagator defined by
\begin{eqnarray}
G_{M^*}(E) =\bra{M^*}\frac{1}{E-H+i\epsilon}\ket{M^*}.
\label{eq:mstar-g0}
\end{eqnarray}
With the projection operator method, as applied in Ref.~\cite{msl},
one can show that Eq.~(\ref{eq:mstar-g0}) in the rest frame of $M^*$ can be written as
\begin{eqnarray}
G^{-1}_{M^*}(E)= E- M^0_{M^*} - \Sigma_{M^*}(E)\,,
\label{eq:mstar-g1}
\end{eqnarray}
where $ M^0_{M^*}$ is a bare mass and
\begin{eqnarray}
\Sigma_{M^*}(E) &=& \sum_{cR,c'R'} \bra{M^*} \Gamma_{M^*,cR}G_{c R,c'R'}(E)
\ket{\bar{\Gamma}_{c'R',M^*}}.
\label{eq:mstar-sigma}
\end{eqnarray}
Here $\ket{\bar{\Gamma}_{c'R',M^*}}$ is defined in Eq.~(\ref{eq:mf-dressed}).

The resonance pole positions, $E_{\text{pole}}$, are defined as zeros of $G^{-1}_{M^*}(E)$. 
They are on the unphysical sheets of the complex-energy Riemann surface and 
are thus defined by the following equation:
\begin{eqnarray}
G^{-1}_{M^*}(E_{\text{pole}}) = 0 .
\label{eq:pole}
\end{eqnarray}
We use the analytic continuation method of Refs.~\cite{ssl1,ssl2} to solve
Eq.~(\ref{eq:pole}) and find $E_{\text{pole}}$ for the considered
coupled-channels model with unstable meson-$R$ channels.

If we replace the dressed vertex $\ket{\bar{\Gamma}_{cR,M^*}}$ by the bare
vertex $\Gamma_{cR,M^*}\ket{M^*}$ in calculating $\Sigma_{M^*}(E_{\text{pole}})$ of 
Eq.~(\ref{eq:mstar-sigma}), then the solution of Eq.~(\ref{eq:pole}) is 
the pole position of the isobar model, which does not include the three-mesons final state interactions.

We now note that the Green function~(\ref{eq:mstar-g1}) can be related
to the excitation of $M^*$ in the $cR \to c'R'$ transition amplitude
$T^{\text{res}}_{c'R',cR}(E)$.
The matrix element of Eq.~(\ref{eq:low}) between $\bra{c'R'}$ and $\ket{cR}$ states is 
\begin{eqnarray*}
T_{c'R',cR}(E) &=& T'_{c'R',cR}(E) + T^{\text{res}}_{c'R',cR}(E) ,
\end{eqnarray*}
where the first term has been defined in Eq.~(\ref{eq:rpi}).
The second term is the ``resonant'' part, and is shown to be
(using the projection operator methods)
\begin{eqnarray}
T^{\text{res}}_{c'R',cR}(E) &=& 
\frac{ \inp{c'R'}{ \bar{\Gamma}_{c'R',M^*}} \inp{\bar{\Gamma}_{M^*,cR}}{cR} }
{E-M^0_{M^*}-\Sigma_{M^*}(E)} ,
\label{eq:t-mstar}
\end{eqnarray}
where
\begin{eqnarray}
\bra{\bar{\Gamma}_{M^*, cR}}
= \sum_{c''R''} 
\bra{M^*} \Gamma_{M^*,c''R''}
\left[\delta_{c''R'',cR} + \sum_{c'''R'''}T'_{c''R'',c'''R'''} G_{c'''R''',cR}(E)\right] .
\label{eq:mf-dressed2}
\end{eqnarray}

The self-energy  $\Sigma_{M^*}(E)$ in Eq.~(\ref{eq:t-mstar}) is defined in Eq.~(\ref{eq:mstar-sigma}),
and can be related to the decay amplitude [Eq.~(\ref{eq:decay-t})] as follows.
For a derivation, consult Appendix~\ref{app:unitarity}.
Consider the ``decay'' width $\Gamma^{M^*}_{\text{tot}}(E)$ of the \textit{bare}
$M^*$ (not the physical resonance state) defined by
\begin{eqnarray}
\Gamma^{M^*}_{\rm tot}(E) &=& 2\pi \sum_{abc} \delta (E-E_a-E_b-E_c) |T_{abc,M^\ast} (E)|^2  .
\label{eq:width-cal} 
\end{eqnarray}
By using the unitarity relation for $T'$ in Eq.~(\ref{eq:low'}), 
we can actually show that the right hand side of the above equation is
\begin{eqnarray}
2\pi \sum_{abc} \delta (E-E_a-E_b-E_c)
|T_{abc,M^\ast} (E)|^2 
&=&-2 {\rm Im} [\Sigma_{M^*}(E)]  .
\label{eq:u-cond1}
\end{eqnarray}
Equation~(\ref{eq:u-cond1}) is used to check the accuracy of our numerical calculations
of the Dalitz plots that are calculated from $T_{abc,M^\ast} (E)$
using the formula detailed in Appendix~\ref{app:dalitz}.

\section{\label{sec:nmr}Formula for numerical calculations}

For numerical calculations of decays of $M^\ast$ into three light pseudoscalar mesons, 
$M^\ast\to abc$ (Fig.~\ref{fig:mstar-decay}), 
we perform  partial-wave expansions of the equations presented in Sec.~\ref{sec:frm} 
in the $M^\ast$ rest frame.
The kinematics of this decay is specified by the following:
\begin{eqnarray}
M^\ast (\vec{0},S_{M^\ast}^z,T_{M^\ast}^z) &\to&
R(\vec{p}_R,s^z_{R},t^z_{R}) + c (\vec{p}_c,0,t^z_{c}) 
\nonumber\\
&\to& 
a (\vec{p}_a,0,t^z_{a})  + b (\vec{p}_b,0,t^z_{b}) + c (\vec{p}_c,0,t^z_{c}) ,
\end{eqnarray}
where the variables in the parenthesis for each particle are its momentum, $z$ components
of the spin and isospin, respectively.

Our task in this section is to relate the  partial-wave forms
of all of the equations presented in Sec.~\ref{sec:frm} to the basic input that
will be  determined by using the available empirical meson-meson 
scattering ($\pi\pi\to\pi\pi$, $\pi K\to\pi K$, etc) amplitudes. 
In this way, the final three-meson scattering effects can be predicted 
for investigating heavy-meson decays.

\subsection{$R\to ab$ decays}

In a decay $R\to ab$, 
the spin ($s_R$) [isospin ($t_R$)] of the parent $R$ state is the same as 
the relative orbital angular momentum $L_{ab}$ [total isospin $I_{ab}$]
of the two-body $ab$ system.
Thus, the partial-wave expansion of the vertex
function $f_{ab,R}$ in the rest frame of $R$ is
\begin{eqnarray}
f_{ab,R} (\vec{q}) &=&
\inp{t_a\, t^z_{a}\, t_b\, t^z_{b}}{t_R\, t^z_{R}} Y_{s_R,s^z_R}(\hat q)\,
\tilde{f}_{ab,R}^{L_{ab}I_{ab}}(q) .
\label{eq:R-ab-cm}
\end{eqnarray}
Here $t_a$ is the isospin of meson $a$ and $t^z_a$ is its $z$-component;
$\inp{j_1 m_1 j_2 m_2}{JM}$ is the Clebsch-Gordan coefficient;
$\vec{q}$ is the relative momentum between $a$ and $b$;
$\tilde{f}_{ab,R}^{L_{ab}I_{ab}}(q)$ is a scalar function
satisfying $\tilde{f}_{ab,R}^{L_{ab} I_{ab}}(q)=0$ for 
$L_{ab}\not= s_R$ and/or $I_{ab}\not= t_R$.

We use the parametrization
\begin{eqnarray}
\tilde{f}^{L_{ab} I_{ab}}_{ab,R}(q)=
\delta_{s_R,L_{ab}}\delta_{t_R,I_{ab}}\frac{g_{ab,R}}{\sqrt{m_\pi}}
\left[\frac{1}{1+(q/c_{ab,R})^2}\right]^{1+(L_{ab}/2)}
\left( \frac{q}{m_\pi} \right)^{L_{ab}}  .
\label{eq:pipi-vertex}
\end{eqnarray}
The parameters $g_{ab,R}$ and $c_{ab,R}$ and the bare mass $m_R$ of $R$ are adjusted
to fit the empirical partial-wave amplitudes. 
The number of bare $R$ states included in the model depends on 
a partial wave considered and the energy region covered in the fit.

\subsection{The $\pi\pi$ model}

We give an expression for the amplitudes of the scattering of two light pseudoscalar mesons in the
partial-wave basis. Here, we limit ourselves to only $\pi\pi$ scattering 
because we consider only three-pion heavy-meson decays in this work.
To fit $\pi\pi$ data up to invariant mass $W=2$ GeV, 
we include $\pi\pi$ and $K\bar{K}$ channels.
Then Eq.~(\ref{eq:pipit}) with total angular momentum $L$ and 
total isospin $I$ in each partial-wave is of the following analytic form
[note that $\tilde f^{L_{ab}I_{ab}}_{R,ab}(q)= \tilde f^{L_{ab}I_{ab}\ast}_{ab,R}(q)$]:
\begin{eqnarray}
T^{LI}_{\pi\pi,\pi\pi} (q',q; E) =
\sum_{R',R}
\bar{f}^{LI}_{\pi\pi,R'}(q') \tau^{LI}_{R',R}(E) \bar{f}^{LI}_{R,\pi\pi}(q) ,
\label{eq:pw-pipi-t}
\end{eqnarray}
with
\begin{eqnarray}
[(\tau^{LI})^{-1}(E)]_{R'R} = (E-m_R)\delta_{R',R} -\Sigma^{LI}_{R',R}(E) ,
\end{eqnarray}
where $m_R$ is the bare mass of $R$ and
\begin{eqnarray}
\Sigma^{LI}_{R',R}(E) = \sum_{ab=\pi\pi,K\bar{K}}
\int_0^\infty q^2 dq 
\frac{\bar{f}^{LI}_{R',ab}(q) \bar{f}^{LI}_{ab,R}(q)} {E - E_a(q) - E_b(q) + i\epsilon} ,
\label{eq:pipi-seigma}
\end{eqnarray}
and
\begin{equation}
\bar{f}^{L_{ab},I_{ab}}_{ab,R}(q) = 
\left\{
\begin{array}{ll}
\displaystyle
\frac{1}{\sqrt{2}}\tilde{f}^{L_{ab},I_{ab}}_{ab,R}(q) & \text{(if $a$ and $b$ are identical particles)}, \\
\displaystyle
\tilde{f}^{L_{ab},I_{ab}}_{ab,R}(q) & \text{(otherwise)} .
\end{array}
\right.
\label{eq:tilde-f}
\end{equation}

\subsection{Coupled-channels equations for $cR\to c'R'$ scattering}

For given total angular momentum $J$, parity $P$, and total isospin $T$, 
the partial-wave form of Eq.~(\ref{eq:rpi}) for the $cR\to c'R'$ scattering can be written as
\begin{eqnarray}
T'^{JPT}_{(c'R')_{l'},(cR)_{l}} (p',p; E)
&=&
Z^{JPT}_{(c'R')_{l'},(cR)_{l}} (p',p; E)
\nonumber\\
&&
+ \sum_{(c'''R''')_{l'''},(c''R'')_{l''}}
\int^\infty_0 q^2dq Z^{JPT}_{(c'R')_{l'},(c'''R''')_{l'''}}(p',q;E) 
G_{(c'''R''')_{l'''},(c''R'')_{l''}}(q,E) 
\nonumber\\
&&
\qquad \times
T'^{JPT}_{(c''R'')_{l''},(cR)_{l}}(q,p;E) .
\label{eq:pw-tcr}
\end{eqnarray}
Here $(cR)_{l}$ denotes the $cR$ state with the relative angular momentum $l$
allowed for given $JPT$;
$p$ ($p'$) is the magnitude of the incoming (outgoing) relative momentum of the $cR$ ($c'R'$) state.
The Green function can be written as 
\begin{eqnarray}
[ G^{-1}(q,E) ]_{(c'''R''')_{l'''},(c''R'')_{l''}}
&=& \delta_{l''',l''}\delta_{c''', c''}
\{
[ E - E_{c''}(q) - E_{R''}(q) ] \delta_{R''',R''} 
\nonumber\\
&&
\qquad \qquad \qquad \qquad
- \Sigma^{c''}_{R''',R''}\left(q, E - E_{c''}(q)\right)
\} ,
\nonumber\\
\label{eq:green-Rc}
\end{eqnarray}
where the self-energy $\Sigma^c_{R',R}(q, w)$ is calculated from Eq.~(\ref{eq:r-selfe})
by inserting the partial-wave expansion~(\ref{eq:R-ab-cm})
and performing a Lorentz transformation to boost the function $\tilde{f}^{L_{ab}I_{ab}}_{ab,R}(q)$
from the rest frame of $R$ to the center-of-mass frame of the $cR$ system. 
We also need to symmetrize the intermediate states with identical mesons. 
Explicitly, we have
\begin{eqnarray}
\Sigma^c_{R'R}(p,E) &=& 
\sum_{ab} \sqrt{\frac{m_{R'}m_{R}}{E_{R'}(p)E_{R}(p)}}
\int_0^\infty q^2 dq
\frac{M_{ab}(q)}{[M^2_{ab}(q) + p^2]^{1/2}}
\frac{ \bar{f}^{L_{ab}I_{ab}}_{R', ab}(q) \bar{f}^{L_{ab}I_{ab}}_{ab,R}(q)}
{E - E_c(p) - [M^2_{ab}(q) + p^2]^{1/2} + i\epsilon} ,
\nonumber\\
\label{eq:R-self}
\end{eqnarray}
where the summation is over all two-mesons states $ab$ of $R\to ab$ decay,
$M_{ab}(q) =  E_{a}(q)+E_{b}(q)$.
The  partial-wave matrix elements $Z^{JPT}_{(c'R')_{l'},(cR)_l}(p',p;E)$ in
Eq.~(\ref{eq:pw-tcr}) of the $Z$-diagram mechanisms, defined by  
Eq.~(\ref{eq:z}), are given in Appendix~\ref{app:z} for the case that
$R$ decays into two pseudoscalar mesons.

\subsection{The $M^*\to abc$ decay amplitudes}

The amplitude [Eq.~(\ref{eq:decay-ampf})] for a strong three-mesons decay is given by
\begin{eqnarray}
\label{eq:decay-amp-amp}
T_{(ab)c, M^\ast} (\vec{p}_a,\vec{p}_b,\vec{p}_c; E) &=&
\sum_{R'R} f_{ab,R'}(\vec{p}_a,\vec{p}_b)  G_{cR',cR}(p_c,E)  \bar{\Gamma}_{cR,M^\ast}(\vec{p}_c,E) ,
\label{eq:decay-amp-um}
\end{eqnarray}
where the Green function $G_{cR',cR}$  has been defined 
by Eqs.~(\ref{eq:green-Rc}) and~(\ref{eq:R-self}).

The $R \to ab$ vertex function in Eq.~(\ref{eq:decay-amp-um}) is obtained from boosting
the matrix element $f^{L_{ab}I_{ab}}_{ab,R}(\vec{q})$, defined by Eq.~(\ref{eq:R-ab-cm})
in the rest frame of $R$, to a moving frame where $\vec{p}_R = \vec{p}_a+\vec{p}_b$:
\begin{eqnarray}
f_{ab,R} (\vec{p}_a,\vec{p}_b) &=&
\sqrt{ \frac{m_{R} E_a(q) E_b(q)}{E_{R}(p_R) E_a(p_a) E_b(p_b)} }
\inp{t_a t^z_{a} t_b t^z_{b}}{t_R t^z_{R}} Y_{s_{R},s^z_{R}}(\hat q)
\tilde{f}^{L_{ab}I_{ab}}_{ab,R}(q) ,
\label{eq:R-ab}
\end{eqnarray}
where $\vec{q}$ is the relative momentum between $a$ and $b$ in their
center-of-mass system; the relation among $\vec{q}$, $\vec{p}_a$ and
$\vec{p}_b$ can be seen in Appendix~\ref{app:z} [Eqs.~(\ref{eq:qa}) and~(\ref{eq:qb})].
For the strong decays, $\bar{\Gamma}_{cR,M^\ast}$ in Eq.~(\ref{eq:decay-amp-um})
for the dressed $M^\ast\to cR$ vertex 
can be written down by using  Eq.~(\ref{eq:mf-dressed}) as
\begin{eqnarray}
\bar{\Gamma}_{cR,M^\ast} (\vec{p}_c,E)
&=&
\sum_{ l, l^z,s^z_R}
\inp{l l^z s_R s^z_R}{S_{M^\ast} S^z_{M^\ast}}
\inp{t_R t^z_R t_c t^z_c}{T_{M^*} T^z_{M^*}}
Y_{l,l^z}(-\hat p_c) \bar F_{(cR)_l,M^\ast}(p_c,E),
\nonumber\\
\label{eq:dressed_mstar}
\end{eqnarray}
where $S_{M^\ast}$ and $T_{M^\ast}$ are the spin and isospin of $M^*$, and
(denoting the parity of $M^\ast$ as $P_{M^\ast}$)
\begin{eqnarray}
\label{eq:dressed-g}
\bar F_{(cR)_l ,M^\ast}(p_c ,E)&=&
F_{(cR)_l , M^\ast}(p_c)
+ \sum_{(c''R'')_{l''},(c'R')_{l'}} \int_0^\infty dq\, q^2 T'^{S_{M^\ast} P_{M^\ast} T_{M^\ast}}_{(cR)_{l},(c''R'')_{l''}}(p_c,q;E) 
\nonumber\\
&&
\qquad\qquad \qquad\qquad \qquad\qquad
\times
G_{(c''R'')_{l''},(c'R')_{l'}}(q,E) F_{(c'R')_{l'},M^\ast}(q) .
\end{eqnarray}
We parametrize the bare vertex function  $F_{(cR)_l,M^\ast}(p)$ as
\begin{eqnarray}
F_{(cR)_l,M^*}(p) = 
\frac{1}{(2\pi)^{3/2}} \frac{C_{(cR)_l,M^*}}{\sqrt{\Lambda_0}}
\left(
\frac{\Lambda^2_{(cR)_l ,M^*}}{p^2 + \Lambda^2_{(cR)_l,M^\ast}}
\right)^{2+(l/2)} 
\left(\frac{p}{m_\pi}\right)^{l} ,
\label{eq:bare_mstar}
\end{eqnarray}
where $C_{(cR)_l,M^*}$, $\Lambda_{(cR)_l,M^*}$ and $m_\pi$ are the
coupling, cutoff and the pion mass, respectively;
$\Lambda_0$ is a scale factor, and is set to be $\Lambda_0 = 1$~GeV.
The couplings $C_{(cR)_l,M^*}$ are nonzero only when the transition
$M^*\to (cR)_l$ is allowed by symmetries, e.g.,
those are nonzero only when $l$ satisfies
$|S_{M^\ast}-s_R|\leq l \leq S_{M^\ast} + s_R$ and $P_{M^\ast}=P_R\times(-)^{l+1}$.
Here it is noted that for a strong decay the bare vertex function $F_{(cR)_l,M^\ast}(p)$ 
is related with $\Gamma_{cR,M^\ast}$ as
\begin{eqnarray}
\Gamma_{cR,M^\ast}(\vec p) = 
\sum_{l,l^z,s_R^z} \inp{t_c t^z_{c} t_R t^z_{R}}{T_{M^\ast}, t^z_R+t^z_c} 
\inp{ll^z s_R s^z_{R}}{S_{M^\ast}, S^z_{M^\ast}} 
Y_{l,l^z}(-\hat p) F_{(cR)_l,M^\ast}(p) .
\end{eqnarray}

For describing the weak decays of $M^*$ such as $D^0$, the above expressions
of  the $M^*\to cR$ vertex function need to be modified to include 
the isospin nonconserving $\Delta T \neq 0$ transition.
This will not be considered here. Instead, we are interested only
in the importance of three-meson scattering after the weak decay of $D^0$,
and it is sufficient to use the above parametrization by extending
$F_{(cR)_l,M^\ast}$ in Eq.~(\ref{eq:bare_mstar}) to depend on $JPT$ of $cR$ state.

The amplitude for the commonly used isobar model, 
as defined by Eq.~(\ref{eq:t-isobar}), is
\begin{eqnarray}
T^{\text{Isobar}}_{(ab)c, M^\ast} (\vec{p}_a,\vec{p}_b,\vec{p}_c; E) 
&=&
\sum_{R'R} f_{ab,R'} (\vec{p}_a,\vec{p}_b) G_{cR',cR}(p_c,E) \Gamma_{cR,M^\ast} (\vec{p}_c) .
\label{eq:decay-amp-isobar}
\end{eqnarray}
We see from Eqs.~(\ref{eq:decay-amp-um}) and~(\ref{eq:decay-amp-isobar}) 
that three-mesons decay amplitudes of these two models differ from each other
only in the functions describing the $M^*\to cR$ decay.
The formulae for calculating the Dalitz plots of final three-meson distributions from these
decay amplitudes [Eqs.~(\ref{eq:decay-amp-um}) and~(\ref{eq:decay-amp-isobar})]
are given in Appendix~\ref{app:dalitz}.

\subsection{\label{sec:nmr-pole}Determinations of resonance positions}

With the partial-wave expansion~(\ref{eq:dressed_mstar}), the resonance pole
condition  $G^{-1}_{M^*}(E_{\text{pole}}) =0 $ of Eq.~(\ref{eq:pole}) leads to
\begin{eqnarray}
E_{\text{pole}}= M^0_{M^*} + \Sigma_{M^*}(E_{\text{pole}}) ,
\label{eq:pole1}
\end{eqnarray}
where [Note that $F_{M^\ast,(c'R')_{l'}}(q)=F^\ast_{(c'R')_{l'},M^\ast}(q)$.]
\begin{eqnarray}
\Sigma_{M^*}(E)=\sum_{(c'R')_{l'},(cR)_{l}}\int_C dq q^2
{F}_{M^*,(c'R')_{l'}}(q)G_{(c'R')_{l'},(cR)_l}(q,E) 
\bar{F}_{(cR)_l,M^*}(q,E) .
\label{eq:mstar-selfep}
\end{eqnarray}
Here $\bar{F}_{(cR)_l,M^*}(q,E)$ has been defined by Eq.~(\ref{eq:dressed-g});
$\displaystyle \int_C dq$ means the momentum integral is performed along
the complex momentum path $C$.

We apply the analytic continuation method developed in  Refs.~\cite{ssl1,ssl2}
to find resonance poles from solving Eqs.~(\ref{eq:pole1}) and~(\ref{eq:mstar-selfep})
for the considered model.

\subsection{Relations of the isobar models with the Breit-Wigner parametrization}

For the isobar model defined within our formulation, we can establish
some relations with the commonly used Breit-Wigner parametrization. 
If we neglect the $Z$-diagram effects by setting
$\bar{F}_{(cR)_l,M^*}$ to $F_{(cR)_l,M^*}$ 
and use the partial-wave expansion~(\ref{eq:dressed_mstar}),  
Eq.~(\ref{eq:t-mstar}) for $cR\to c'R'$ scattering can be written as 
(omitting the momentum variables)
\begin{eqnarray}
T^{\text{res},JPT}_{(c'R')_{l'},(cR)_l}(E) \to  
\frac{F_{(c'R')_{l'},M^*} F_{M^\ast,(cR)_l} }{E - M^0_{M^*} - \Sigma^{(0)}_{M^*}(E)} ,
\label{eq:t-mstar-a}
\end{eqnarray}
where $J=S_{M^\ast}$, $P=P_{M^\ast}$, $T=T_{M^\ast}$ for strong decays, and
\begin{eqnarray}
\Sigma^{(0)}_{M^*}(E) 
=\sum_{(c'R')_{l'},(cR)_{l}}\int_0^\infty q^2 dq
F_{M^*,(c'R')_{l'}}(q)G_{(c'R')_{l'},(cR)_l}(q,E) 
F_{(cR)_l,M^*}(q) .
\label{eq:sigma-0}
\end{eqnarray}
Equation~(\ref{eq:t-mstar-a}) is similar to that of the commonly used Breit-Wigner 
parametrization in the analysis using the isobar model or the $K$-matrix model:
\begin{eqnarray}
T^{\text{BW},JPT}_{(c'R')_{l'},(cR)_l}(E) &=& 
\frac{
[e^{i\delta_{(c'R')_{l'}}}\sqrt{\Gamma^{\text{BW}}_{(c'R')_{l'}}/2}]
[e^{i\delta_{(cR)_l}}\sqrt{\Gamma^{\text{BW}}_{(cR)_l}/2}]}
{E-  M^{\text{BW}}_r +i(\Gamma^{\text{BW}}_{\text{tot}}/2)} ,
\label{eq:t-bw}
\end{eqnarray}
where $\Gamma^{\text{BW}}_{(cR)_l}$ is a partial decay width for $M^\ast\to (cR)_l$, which is related to
the total decay width as
\begin{eqnarray}
\Gamma^{\text{BW}}_{\text{tot}} &=& \sum_{(cR)_l} \Gamma^{\text{BW}}_{(cR)_l} .
\label{eq:width-bw}
\end{eqnarray}
Now we introduce $\bar s_R$ and $\bar t_R$ that specify the spin and isospin of $R$.
Then Eq.~(\ref{eq:sigma-0}) can be written as
\begin{eqnarray}
\Sigma^{(0)}_{M^*}(E) &=& \sum_{\bar s_R,\bar t_R} \left[\Sigma^{(0)}_{M^*}(E)\right]_{\bar s_R,\bar t_R} ,
\end{eqnarray}
where
\begin{eqnarray}
\left[\Sigma^{(0)}_{M^*}(E)\right]_{\bar s_R,\bar t_R} &=&
\sum_{\{(c'R')_{l'} | s_{R'}=\bar s_R, t_{R'}=\bar t_R\}}
\sum_{\{(cR)_l | s_R =\bar s_R, t_R= \bar t_R\}}
\nonumber\\
&&
\qquad\qquad
\times
\int_0^\infty q^2 dq F_{M^*, (c'R')_{l'}}(q) G_{(c'R')_{l'},(cR)_{l}}(q,E)  F_{{(cR)_l},M^*}(q) .
\label{eq:part-sigma}
\end{eqnarray}
We denote a conditional sum of $(cR)_l$ by
$\sum_{\{(cR)_l | s_R=\bar s_R, t_R=\bar t_R\}}$,
in which the $(cR)_l$ state is summed, keeping $s_R$ and $t_R$ constant,
i.e., $s_R=\bar s_R$ and $t_R=\bar t_R$.
Thus it is reasonable to make the  following interpretations
\begin{eqnarray}
M^{\text{BW}}_r &=& M^0_{M^*} + \mathrm{Re} \left[ \Sigma^{(0)}_{M^*}(M^{\text{BW}}_r)\right] ,
\label{eq:m-bw-um} \\
\Gamma^{\text{BW}}_{\text{tot}}(M^{\text{BW}}_r) &=& 
-2 \mathrm{Im}\left[\Sigma^{(0)}_{M^*}(M^{\text{BW}}_r)\right]  ,
\label{eq:w-bw-um} \\
\Gamma^{\text{BW}}_{\bar s_R, \bar t_R}(M^{\text{BW}}_r) &=& 
-2 \mathrm{Im}\left[[\Sigma^{(0)}_{M^*}(M^{\text{BW}}_r)]_{\bar s_R, \bar t_R}\right] .
\label{eq:part-w-bw-um}
\end{eqnarray}
Equations~(\ref{eq:m-bw-um})-(\ref{eq:part-w-bw-um}) will be used in our later comparisons with
the data listed by PDG. 
We note here that the above identifications are very qualitative.

\section{\label{sec:appl}Application}

In this section, we apply our model explained in the previous sections to
investigate the three-pions decays of heavy mesons
$a_1(1260)$, $\pi_2(1670)$, $\pi_2(2100)$, and also $D^0$.
Our first task is to determine the parameters of our model.
To simplify the calculations, we determine the vertex interactions
$R\rightarrow ab$ for $ab=\pi\pi, K\bar{K}$ by fitting only the $\pi\pi$ 
scattering phase shifts up to the invariant mass $W =2000 $ MeV.
This is clearly a simplification since the data associated with $K\bar{K}$
channel should in principle be included in our fits and we must also include
four-pions channels that have been considered to
be important in the isoscalar-scalar ($L=I=0$) partial wave.
However, such a detailed study of meson-meson scattering can only be done 
rigorously by extending our formulation to account for the direct
meson-meson interactions $v_{a'b',ab}$ which must be carefully derived from
effective field theory approaches, e.g., Refs.~\cite{gas85,hls88,eck89,ber91},
to make sure that the predicted
$\pi\pi$ amplitudes near threshold have the analytic properties
constrained by the chiral symmetry. 
Furthermore, the inclusion of $v_{a'b',ab}$ in our model Hamiltonian $H^{''}$ of 
Eq.~(\ref{eq:h-int2}) will greatly complicate the scattering formulation, 
as can be seen in the $\pi \pi N$ formulation presented in Ref.~\cite{msl}.
For our present limited purpose of demonstrating the importance
of three-body unitarity, our simplified model that reproduces $\pi\pi$
phase shifts in $s$, $p$ and $d$ waves up to $W=2000$ MeV is sufficient.
For the same reasons, we neither include $\pi K\bar K$ $Z$-diagrams nor
make an attempt to estimate the errors of the determined parameters.

In Sec.~\ref{sec:appl-par}, we determine the model parameters by fitting
the $\pi\pi$ phase shifts and resonance parameters listed in PDG.
With the parameters obtained from the fit,
we determine the pole positions (Sec.~\ref{sec:pole})
and calculate the Dalitz plots (Sec.~\ref{sec:dalitz})
from the $M^*\to \pi\pi\pi$  amplitudes
including the $Z$-diagram [Eq.~(\ref{eq:decay-amp-um})]
or without the $Z$-diagram [Eq.~(\ref{eq:decay-amp-isobar})]
with the formula given in Appendix~\ref{app:unitarity}.
For the calculation without the $Z$-diagram,
we either simply turn off the $Z$-diagram in the full calculation,
or fit the isobar model to the Dalitz plot from the full calculation.
Our main focus is to examine the effect of the $Z$-diagram 
(and thus the three-body unitarity) on these quantities
by detailed comparison of the results calculated 
with and without $Z$-diagram mechanisms,
thereby providing information about the extent to which 
the commonly used isobar-model analysis is valid for
extracting the properties of heavy meson from three-mesons decay data.

\subsection{\label{sec:appl-par}Determinations of model parameters}

\subsubsection{Fits to $\pi\pi$ amplitudes}

Our first task is to determine the $R\to ab$ vertex function
$\tilde{f}^{L_{ab}I_{ab}}_{ab,R}(q)$, defined by
Eq.~(\ref{eq:R-ab-cm}), by fitting the $\pi\pi$ phase shifts.
We include $\pi\pi$ and $K\bar{K}$ channels and use 
the formulae~(\ref{eq:pipi-vertex}) and~(\ref{eq:pw-pipi-t}) 
to fit the available $\pi\pi$ amplitude in $s$, $p$ and $d$ partial waves. 
In our fits, the number of bare $R$ included in each partial wave is 2, 2, 1 for
$(s_R,t_R)= (L_{ab},I_{ab})=(0,0), (1,1), (2,0)$, respectively.
The resulting parameters are listed in Table~\ref{tab:pipi}.

\begin{table}[t]
\caption{\label{tab:pipi} 
Masses ($M_{R_i}$), couplings ($g_{\pi\pi,R_i}$, $g_{K\bar K,R_i}$), 
and cutoffs ($c_{\pi\pi,R_i}$, $c_{K\bar K,R_i}$) of the $i$-th bare $R$ states, $R_i$, 
in the $\pi\pi$ partial wave with the angular momentum $L$ and the isospin $I$.
The couplings and cutoffs are defined in Eq.~(\ref{eq:pipi-vertex}).
}
\begin{ruledtabular}
\begin{tabular}{lcccccccccc}
$R~(L,I)$
& $M_{R_1}$ & $g_{\pi\pi,R_1}$ & $c_{\pi\pi,R_1}$ & $g_{K\bar{K},R_1}$ & $c_{K\bar{K},R_1}$  
& $M_{R_2}$ & $g_{\pi\pi,R_2}$ & $c_{\pi\pi,R_2}$ & $g_{K\bar{K},R_2}$ & $c_{K\bar{K},R_2}$  \\
&(MeV)&&(MeV)&&(MeV)&(MeV)&&(MeV)&&(MeV)\\
\hline
$f_0$ (0, 0) & 1220   & $-$0.898 & 441 & 0.006 & 1970   & 2400 & 0.700 & \,~955 & $-$1.179  & 394 \\
$\rho$~~(1, 1) & \,~891 & $-$0.291 & 394 & 0.106 & \,~467 & 1840 & 0.021 & 1973   & ~~\,0.167 & 394 \\
$f_2$ (2, 0) & 1607   & $-$0.051 & 567 & 0.015 & \,~818 & ---  & ---   & ---    & ---       & --- \\
\end{tabular}
\end{ruledtabular}
\end{table}

\begin{figure}[t]
\includegraphics[width=0.3\textwidth]{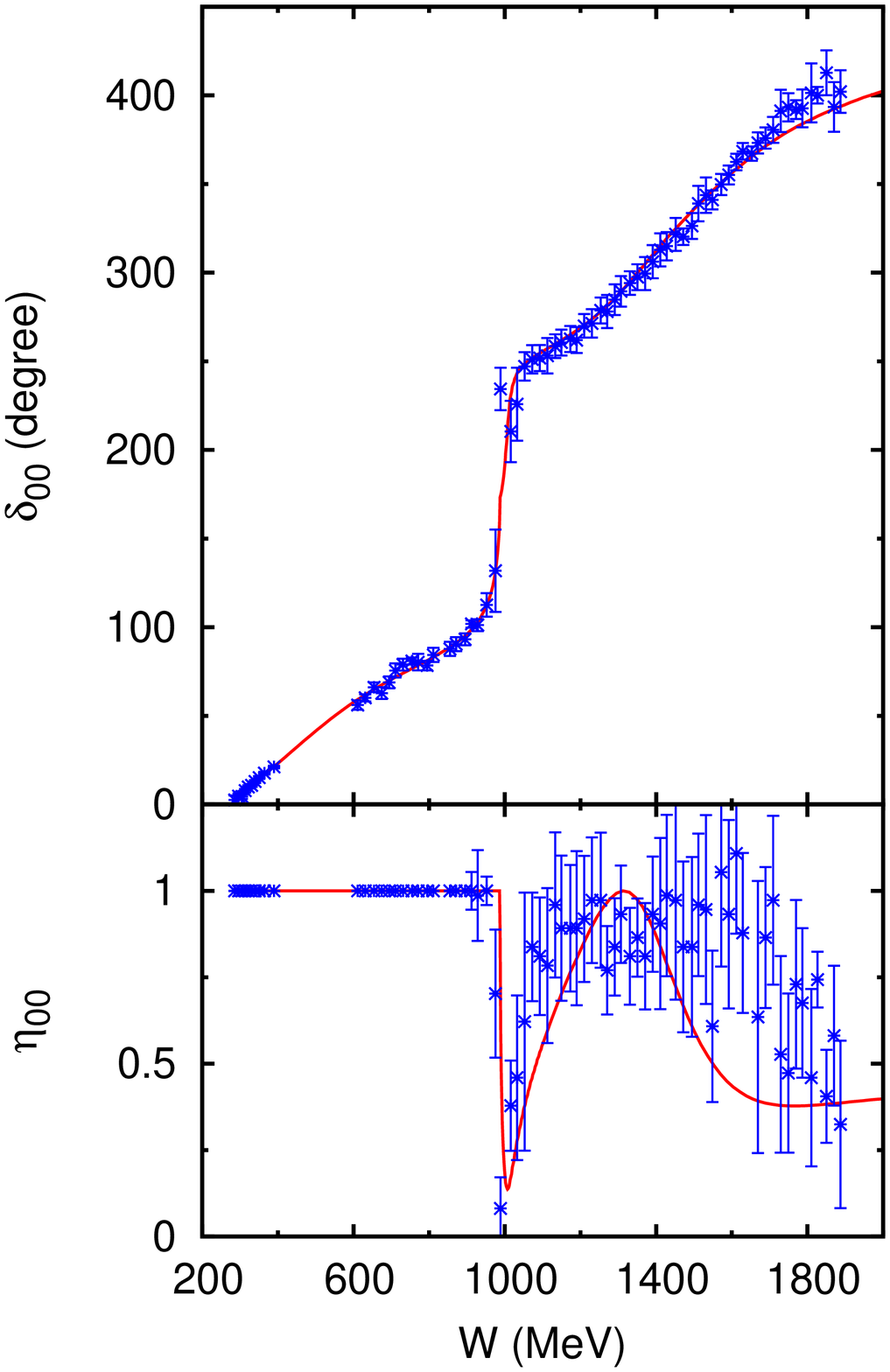}
\includegraphics[width=0.3\textwidth]{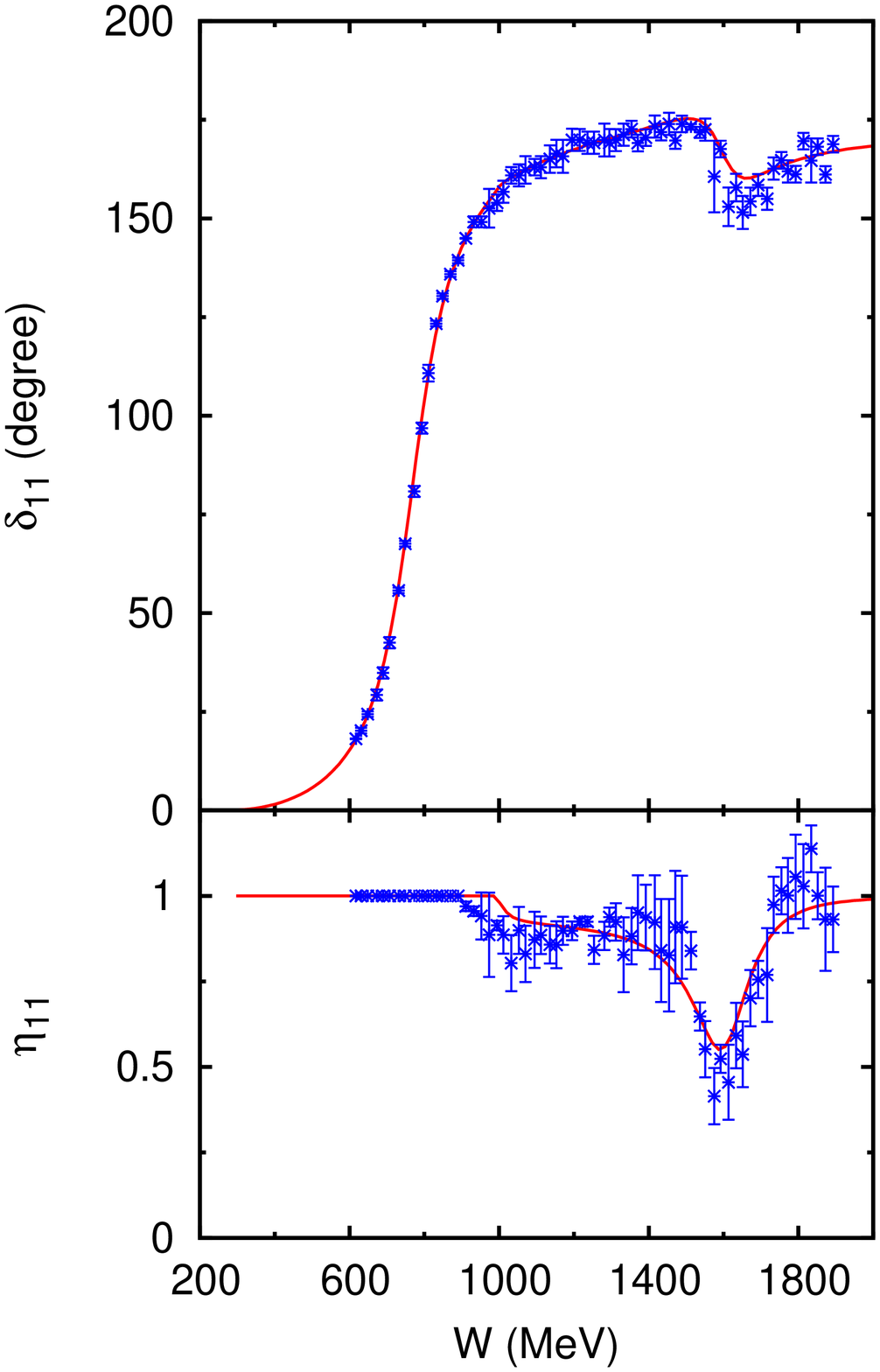}
\includegraphics[width=0.3\textwidth]{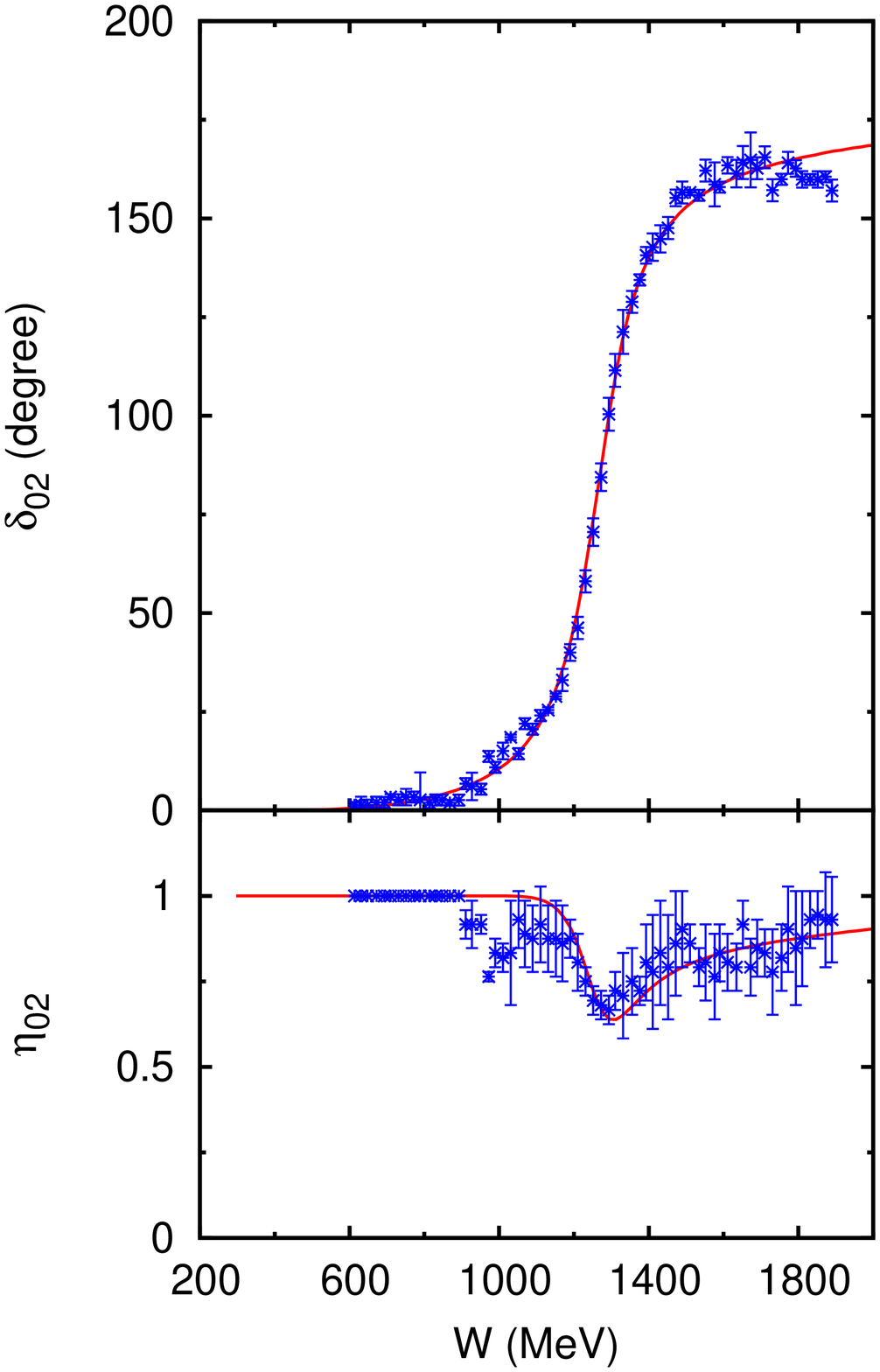}
\caption{\label{fig:pipi} 
(Color online) 
Phase shifts (upper) and inelasticities of the $\pi\pi$ scattering (lower):
(left panels) $L=I=0$, (center panels) $L=I=1$, and (right panels) $L=2,I=0$. 
Data are taken from Ref.~\cite{grayer,hyams,na48}.}
\end{figure}

As shown in Fig.~\ref{fig:pipi}, we are able to get good fits to the
empirical partial-wave amplitudes~\cite{grayer,hyams,na48}.
The nonzero values of the inelasticities are due to the couplings to $K\bar{K}$ channels. 
For $s$ and $p$ waves, $(L_{ab},I_{ab})=(0,0), (1,1)$, the high quality
fits are obtained only when two bare $R$ states are included.
It is noted that a partial-wave analysis using more recent data~\cite{becker} 
has found a unique solution for $W \sim 1000$-$1800$ MeV.
Although our present model is reasonable enough to address the question on the importance 
of the $Z$-graphs, those data should be considered for a more quantitative 
application of our model.

We have also determined the resonance pole positions 
by applying the analytic continuation method of Refs.~\cite{ssl1,ssl2}.
The results for ${\rm Re} \left[E\right]\le 2$~GeV are listed in Table~\ref{tab:pipi-pole}. 
It is interesting to note that we have two bare $R$ states in $s$ wave, 
but we have identified three resonance poles on different sheets of
Riemann surface: sheet II is $(up)$ consisting of the
unphysical ($u$) $\pi\pi$ and physical ($p$) $K\bar{K}$ sheet,
and sheet III is ($uu$).

\begin{table}[t]
\caption{\label{tab:pipi-pole}
Pole positions of the $\pi\pi$ partial-wave amplitudes with the angular momentum $L$ and the isospin $I$
in the complex-energy plane. 
We list only the poles below $\text{Re}[E]\le 2$~GeV.
Roman numerals in the square brackets specify the Riemann sheet on which the pole exists.
We use the convention for specifying each Riemann sheet, I -- IV, which is defined 
in, e.g., Ref.~\cite{sheet}.
}
\begin{ruledtabular}
\begin{tabular}{ccccc}
$L$ & $I$ & \multicolumn{3}{c}{Pole positions (GeV) [Riemann sheet]}  \\\hline
0   & 0   & $0.43 - 0.27i$~\, [II]  & $1.00 - 0.009i$   [II] & $1.35 - 0.17i$  [III] \\
1   & 1   & $0.77 - 0.081i$   [II]  & $1.61 - 0.12i$\, [III] & --- \\ 
2   & 0   & $1.25 - 0.10i$\,  [III] & --- & --- 
\end{tabular}
\end{ruledtabular}
\end{table}

We find that the poles listed in Table~\ref{tab:pipi-pole}
can be identified with  the $\pi\pi$ resonances listed by 
PDG~\cite{pdg2010}. For the $(L,I)=(0,0)$
$s$-wave partial wave, our results can be identified with
$f_0(600)$ (or $\sigma$), $f_0(980)$ and $f_0(1370)$. 
For $(L,I)=(1,1)$ $p$ wave, our results correspond to the 
$\rho(770)$ and a higher mass $\rho$.
The  resonance $f_2(1270)$ can be identified with
our result for the $(L,I)=(0,2)$ $d$-wave partial wave.
Here we note that the imaginary part of the position $(1.00-0.009i)$ GeV,
which corresponds to having 18 MeV of the full width,
in the isoscalar-scalar $L=I=0$ partial wave (the second row of Table~\ref{tab:pipi-pole})
is too small compared with the full width $40 - 100$ MeV of $f_0(980)$ listed by PDG. 
This perhaps can be improved only by extending our model to include
four-pions channel and direct interactions $v_{ab,a'b'}$ with $ab, a'b' = \pi\pi, K\bar{K}$.
But this is beyond the scope of this investigation, as discussed in the 
beginning of Sec.~\ref{sec:appl}.
Here we mention that the imaginary parts of the pole position of $f_0(980)$ 
from some previous $\pi\pi$ models are also smaller than the full width,
(40-100) MeV, listed by PDG,
such as 28 MeV from Ref.~\cite{oll99}, and 29 MeV from Ref.~\cite{klp94}.
A quark model~\cite{wi82} also gave only 15 MeV for the width of $f_0(980)$.

In most of the previous studies of heavy-meson decays, only the
$s$-wave resonances are included as resonance poles
while the $p$-wave poles are included in Ref.~\cite{guo}. 
The other resonances are included using the Breit-Wigner form.
In our calculations, we include all resonance poles in $\pi\pi$
$s$, $p$ and $d$ partial waves listed in Table~\ref{tab:pipi-pole}.

We evaluate the $\pi R$ Green functions [Eq.~(\ref{eq:green-Rc})]
and the matrix elements of $Z$-diagrams [Eq.~(\ref{eq:zj-final})] with
the parameters listed in Table~\ref{tab:pipi}.
We solve the coupled-channels equations~[Eq.~(\ref{eq:pw-tcr})]
to obtain the $\pi R \to \pi' R'$ scattering amplitude 
for given $JPT$, $T^{JPT}_{(\pi' R')_{l'}, (\pi R)_l}(p',p,E)$,
including all allowed relative orbital angular momentum between $\pi$ and $R$.
The resulting $T^{JPT}_{(\pi' R')_{l'}, (\pi R)_l}(p',p,E)$ are then used to 
calculate the $M^*\to \pi\pi\pi$ decay amplitudes~(\ref{eq:decay-amp-um}) 
and find resonance poles associated
with $M^*$ by solving Eqs.~(\ref{eq:pole1}) and~(\ref{eq:mstar-selfep}).

\subsubsection{Parameters for the decays of $M^*$ states}
\label{sec:par-mdecay}

To calculate the decay amplitudes for $a_1(1260)$, $\pi_2(1670)$, $\pi_2(2100)$ and $D^0$, 
we now need to determine their bare masses $M^0_{M^*}$, and the parameters 
$C_{(\pi R)_l,M^*}$, $\Lambda_{(\pi R)_l,M^*}$ of Eq.~(\ref{eq:bare_mstar}) for the 
$M^*\to \pi R$ vertex functions.
Ideally, we should determine these parameters by fitting the
Dalitz plots of $\pi\pi\pi$  distributions measured experimentally.
However, such a rather complex process is not needed for our limited purpose here 
to mainly investigate the extent to which the commonly used isobar-model analysis is valid.
It is sufficient to choose our parameters guided by  the resonance positions and 
branching ratios listed by PDG~\cite{pdg2010}\footnote{
PDG lists small but non-zero branching ratios of decay channels that we do not consider in
our model; we ignore them.}.
The data for the resonances considered in this work are listed in Table~\ref{tab:pi2}.

\begin{table}[t]
\caption{\label{tab:pi2}
Properties of $M^\ast= a_1 (1260),~\pi_2 (1670),~\pi_2 (2100)$ to which our model is fitted:
Isospin ($I$), spin ($J$), parity ($P$), and charge conjugation parity ($C$);
pole masses ; branching ratios (BR).
}
\begin{ruledtabular}
\begin{tabular}{lccc}
            &$a_1 (1260)$&$\pi_2 (1670)$ &$\pi_2 (2100)$ \\ \hline
$I(J^{PC})$ & $1(1^{++})$& $1(2^{-+})$&$1(2^{-+})$ \\
Pole masses (MeV)& $1230 - 213i$& $1672 - 130i$ & $2090 - 313i$ \\
$\text{BR}(M^\ast\to \pi f_0)$ (\%)\hspace*{10mm}&23 & 13 & 45 \\
$\text{BR}(M^\ast\to \pi \rho)$ ~(\%) &74 & 31  & 19 \\
$\text{BR}(M^\ast\to \pi f_2)$ (\%)&~~~\,2.5 &  56  & 35 
\end{tabular}
\end{ruledtabular}
\end{table}

We first notice that the data in Table~\ref{tab:pi2} are the averaged values from several analyses. 
Most of these analyses parametrized the $M^\ast$ decay amplitudes with the Breit-Wigner form, 
and all of them treated the final three pions as the paired two pions
(whose correlations are described by either the Breit-Wigner form or the $K$-matrix)
and the noninteracting spectator.
We thus assume that the $\Gamma^{\text{BW}}_{\text{tot}}/2$ is the imaginary part of 
the pole masses in Table~\ref{tab:pi2}, from which we can use the listed branching ratios
$\text{BR}(M^\ast\to cR) $ to calculate 
$\Gamma^{\text{BW}}_{cR}=\text{BR}(M^\ast \to cR)\times\Gamma^{\text{BW}}_{\text{tot}}$.\footnote{
In data analyses with the Breit-Wigner parametrization of the $M^*$ decay amplitudes,
the partial width and the imaginary part of the $M^\ast$ pole masses are not related by
$\Gamma^{\text{BW}}_{cR}=\text{BR}(M^\ast \to cR)\times\Gamma^{\text{BW}}_{\text{tot}}$.
We use this relation just for determining the parameters with this rough estimate.
}
The resulting values of $\Gamma^{\text{BW}}_{cR}$ as well as
the pole masses of $M^\ast$ are then used as data  
to determine  the parameters of $F_{(cR)_l,M^*}(p)$ [Eq.~(\ref{eq:bare_mstar})]
and $M^0_{M^\ast}$ [Eq.~(\ref{eq:mstar-g1})]
by using Eqs.~(\ref{eq:pole1}), (\ref{eq:w-bw-um}) and~(\ref{eq:part-w-bw-um}).
Because we have more parameters than the number of data,
we use for simplicity the same cutoff for all of $F_{(cR)_l,M^*}(p)$ for a given $M^\ast$, 
and we only adjust the coupling constant $C_{(cR)_{l\text{min}},M^\ast}$ with the lowest 
allowed angular momentum $l_{\text{min}}$ for $cR$;
the other $C_{(cR)_l,M^*}$ are set to zero.
The resulting parameters for $a_1(1260)$, $\pi_2(1670)$, and $\pi_2(2100)$
are listed in Tables~\ref{tab:mstar-par}-\ref{tab:mstar-par-pi22100}.

\begin{table}[t]
\caption{\label{tab:mstar-par} 
Masses ($M^0_{M^\ast}$), cutoffs ($\Lambda_{\pi R,M^\ast}$),
and couplings ($C_{(\pi R^{LI}_i)_l,M^*}$) of the bare $M^\ast = a_1(1260)$.
The cutoffs and couplings are defined in Eq.~(\ref{eq:bare_mstar}). 
For $C_{(\pi R^{LI}_i)_l,M^*}$, $R^{LI}_i$ means the $i$-th bare $R$ state
with the spin $L$ and the isospin $I$, and $l$ denotes the orbital angular momentum
between $R^{LI}_i$ and $\pi$.
The second (third) column shows the parameters for 
the unitary (isobar-fit) model.
The hyphens (---) indicate the unused parameters.
See the text for the definition of the isobar-fit model.
}
\begin{ruledtabular}
\begin{tabular}{lcc}
&\multicolumn{2}{c}{$a_1(1260)$} \\\cline{2-3}
  &Unitary model  & Isobar-fit model  \\\hline
$M^0_{M^*}$ (MeV)          &   1687  &    1901 \\
$\Lambda_{\pi R,M^*}$ (MeV)&   ~\,832   &    1073 \\
$C_{(\pi R^{00}_1)_1,M^*}$ &   \qquad~\,4.46  & \qquad~~2.84 \\
$C_{(\pi R^{00}_2)_1,M^*}$ &  \qquad\,$-$3.41& \quad~~$-$0.13 \\
$C_{(\pi R^{11}_1)_0,M^*}$ &  \qquad 16.8  & \quad~~13.3 \\
$C_{(\pi R^{11}_1)_2,M^*}$ &  ---    & \qquad~~0.15 \\
$C_{(\pi R^{11}_2)_0,M^*}$ &  \qquad\,$-$0.76& ~~\,$-$10.0 \\
$C_{(\pi R^{11}_2)_2,M^*}$ &  ---    & \quad~~$-$0.17 \\
$C_{(\pi R^{20}_1)_1,M^*}$ &  \qquad 10.4  & \qquad~~7.37 \\
$C_{(\pi R^{20}_1)_3,M^*}$ &  ---    & \quad~~$-$0.06 
\end{tabular}
\end{ruledtabular}
\end{table}
\begin{table}[t]
\caption{\label{tab:mstar-par-pi21670} 
Masses ($M^0_{M^\ast}$), cutoffs ($\Lambda_{\pi R,M^\ast}$),
and couplings ($C_{(\pi R^{LI}_i)_l,M^*}$) of the bare $M^\ast = \pi_2(1670)$.
For the description of the table, see the caption of Table.~\ref{tab:mstar-par}.
}
\begin{ruledtabular}
\begin{tabular}{lcc}
  &\multicolumn{2}{c}{$\pi_2(1670)$} \\ \cline{2-3}
  &Unitary model  & Isobar-fit model \\ \hline
$M^0_{M^*}$ (MeV)          &     1877    &    1912  \\
$\Lambda_{\pi R,M^*}$ (MeV)&    ~\,874     &   ~\,885     \\
$C_{(\pi R^{00}_1)_2,M^*}$ &    \qquad~\,0.67     &   \qquad~\,0.55      \\
$C_{(\pi R^{00}_2)_2,M^*}$ &    \qquad~\,0.99     &   \qquad~\,0.97      \\
$C_{(\pi R^{11}_1)_1,M^*}$ &   \quad~~$-$2.21   &   \quad~~$-$1.67   \\
$C_{(\pi R^{11}_1)_3,M^*}$ &      ---    &    ---      \\
$C_{(\pi R^{11}_2)_1,M^*}$ &    \qquad~\,0.50     &    \qquad~\,3.58     \\
$C_{(\pi R^{11}_2)_3,M^*}$ &      ---    &    ---      \\
$C_{(\pi R^{20}_1)_0,M^*}$ &   ~~\,$-$12.2   &   ~~\,$-$11.3   \\
$C_{(\pi R^{20}_1)_2,M^*}$ &      ---    &    ---      \\ 
$C_{(\pi R^{20}_1)_4,M^*}$ &      ---    &    ---   
\end{tabular}
\end{ruledtabular}
\end{table}
\begin{table}[t]
\caption{\label{tab:mstar-par-pi22100} 
Masses ($M^0_{M^\ast}$), cutoffs ($\Lambda_{\pi R,M^\ast}$),
and couplings ($C_{(\pi R^{LI}_i)_l,M^*}$) of the bare $M^\ast = \pi_2(2100)$.
For the description of the table, see the caption of Table.~\ref{tab:mstar-par}.
}
\begin{ruledtabular}
\begin{tabular}{lcc}
  &\multicolumn{2}{c}{$\pi_2(2100)$} \\ \cline{2-3}
  &Unitary model  & Isobar-fit model  \\ \hline
$M^0_{M^*}$ (MeV)         &     2189    & 2280    \\
$\Lambda_{\pi R,M^*}$ (MeV)&     ~\,876     & ~\,970     \\
$C_{(\pi R^{00}_1)_2,M^*}$ &   \quad~~$-$0.93   & \quad~~$-$0.70    \\
$C_{(\pi R^{00}_2)_2,M^*}$ &   \qquad~$-$0.001  & \quad~~$-$0.34    \\
$C_{(\pi R^{11}_1)_1,M^*}$ &     \qquad~\,2.45    & \qquad~\,1.65       \\
$C_{(\pi R^{11}_1)_3,M^*}$ &     ---     &    ---     \\
$C_{(\pi R^{11}_2)_1,M^*}$ &    \qquad~\,0.51     & \qquad~\,0.41       \\
$C_{(\pi R^{11}_2)_3,M^*}$ &     ---     &    ---     \\
$C_{(\pi R^{20}_1)_0,M^*}$ &    ~~\,$-$11.9  & ~~\,$-$10.1    \\
$C_{(\pi R^{20}_1)_2,M^*}$ &     ---     &    ---     \\ 
$C_{(\pi R^{20}_1)_4,M^*}$ &     ---     &    ---     
\end{tabular}
\end{ruledtabular}
\end{table}

The $D^0$ meson (1865 MeV, $J^P = 0^-$) mainly decays weakly and thus three-meson scattering 
effects have very little effect on its mass and width. 
We thus  will only investigate the Dalitz plot for the $D^0\to \pi^+\pi^-\pi^0$ decay.
The BABAR Collaboration~\cite{babar-d1}
presented the Dalitz plot data for this process, and we utilize their observation that
the $D^0\to \pi^+\pi^-\pi^0$ decay is dominated by $T=0$ $\pi\rho$ channel for simplicity.
Thus we use the following parameters: $C_{(\pi R^{11}_1)_1,M^*} = 1$ 
(see Table~\ref{tab:mstar-par} for the notation), and
$C_{(cR)_l,M^*} = 0$ for other partial waves; $\Lambda_{cR,M^*} = 1$~GeV.
We are interested only in the difference between the Dalitz plots calculated with and without
$Z$-diagram, so this simple choice of parameters is sufficient. 
It turns out that this simple choice of the parameters well reproduces
the shape of the Dalitz plot presented by the BABAR Collaboration~\cite{babar-d1}.
Clearly, the above procedure is just for 
a very rough estimate of bare $M^*$ parameters.
In the future, we should fit the Dalitz plot data directly.
But the present procedure is sufficient for our purpose in this paper.

\subsection{$Z$-diagram effects on the pole positions of  $a_1(1260)$, $\pi_2(1670)$, and $\pi_2(2100)$ }
\label{sec:pole}

In Sec.~\ref{sec:par-mdecay}, we solved Eqs.~(\ref{eq:pole1}) and~(\ref{eq:mstar-selfep}) 
to fit the pole positions for $a_1(1260)$, $\pi_2(1670)$ and $\pi_2(2100)$ listed in PDG.
Our fitted values are shown in the row labeled as ``With $Z$'' of Table~\ref{tab:pole}.
When the $Z$-diagram mechanisms are turned off,
which is achieved by replacing the dressed vertex function $\bar{F}_{(cR)_l,M^*}$
with the bare ${F}_{(cR)_l,M^*}$ in calculating $\Sigma_{M^\ast}(E)$ [Eq.~(\ref{eq:mstar-selfep})], 
the solution of Eq.~(\ref{eq:pole1}) becomes the values shown in the row labeled as ``Without $Z$''
of Table~\ref{tab:pole}.

\begin{table}[t]
\caption{\label{tab:pole} 
Pole masses of $a_1(1260)$, $\pi_2(1670)$, and $\pi_2(2100)$.
Here, ``With $Z$'' denotes the results of the full unitary model, while
``Without $Z$'' denotes the results in which the $Z$ diagrams are turned off from the full unitary model.
}
\begin{ruledtabular}
\begin{tabular}{cccc}
          & \multicolumn{3}{c}{Pole masses (MeV)}\\ \cline{2-4}
          & $a_1(1260)$& $\pi_2(1670)$ & $\pi_2(2100)$\\ \hline
With $Z$ & $1230 - 213\ i$ & $1672 - 130\ i$ & $2090 - 313\ i$\\
Without $Z$& $1122 - 148\ i$ & $1661 - 127\ i$ & $2044 - 398\ i$ 
\end{tabular}
\end{ruledtabular}
\end{table}

Comparing the two rows in Table~\ref{tab:pole}, we  see that the $Z$-diagram mechanisms can change
the pole positions significantly. In particular, the imaginary parts can be  changed by
65 MeV for $a_1(1260)$ and 85 MeV for $\pi_2(2100)$
Accordingly, we expect that the extracted residues will also be
significantly changed. The extraction of the residues for unstable particle channels
is nontrivial, and is still being investigated, as explained in Ref.~\cite{ssl2}.
We thus do not have results for the $Z$-diagram effects on the branching ratios in this work.

\subsection{$Z$-diagram effects on Dalitz plots}
\label{sec:dalitz}

\subsubsection{$D^0 \to \pi^+\pi^-\pi^0$}

\begin{figure}[t]
\includegraphics[width=95mm]{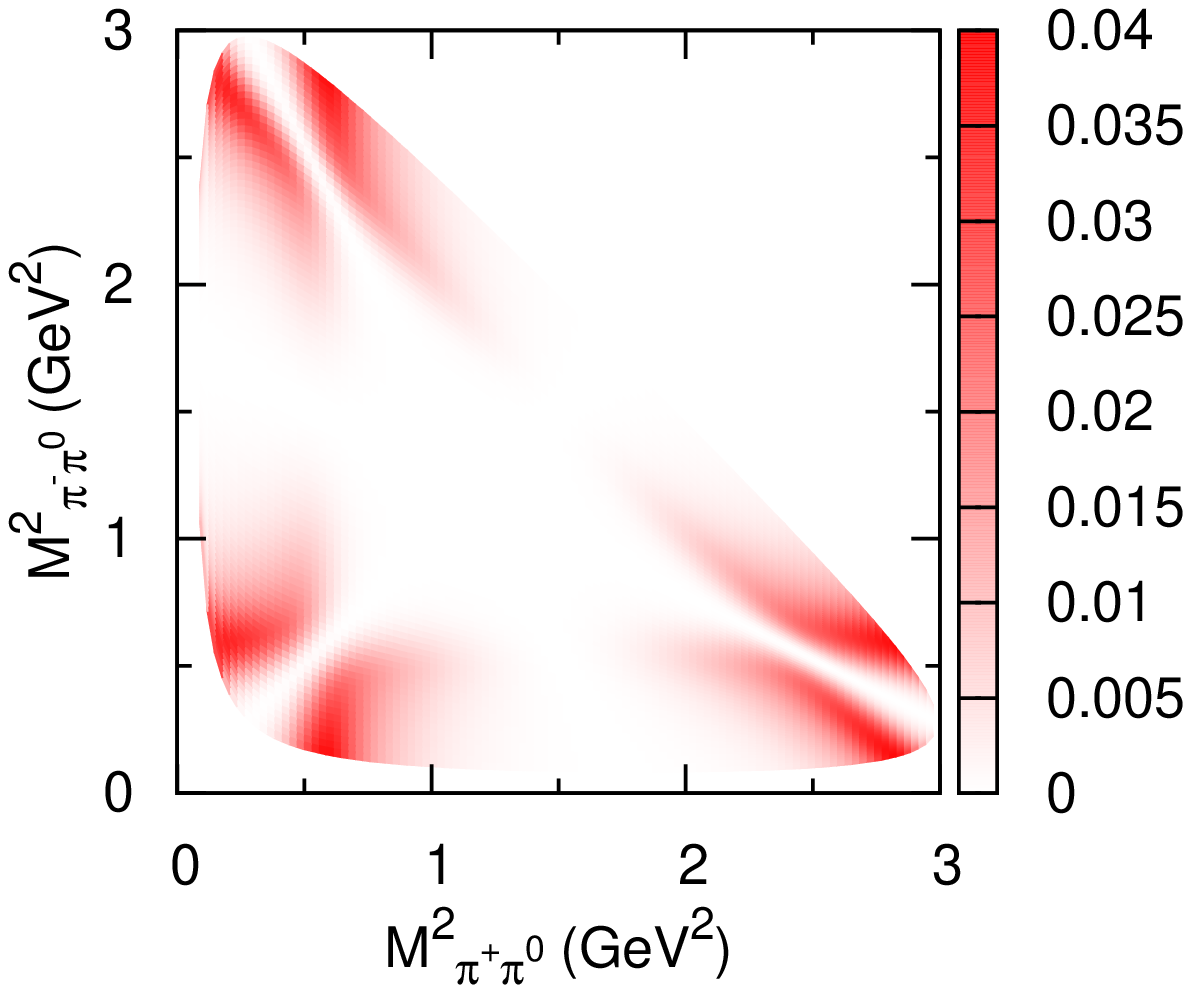}\hspace{-28mm}
\includegraphics[width=95mm]{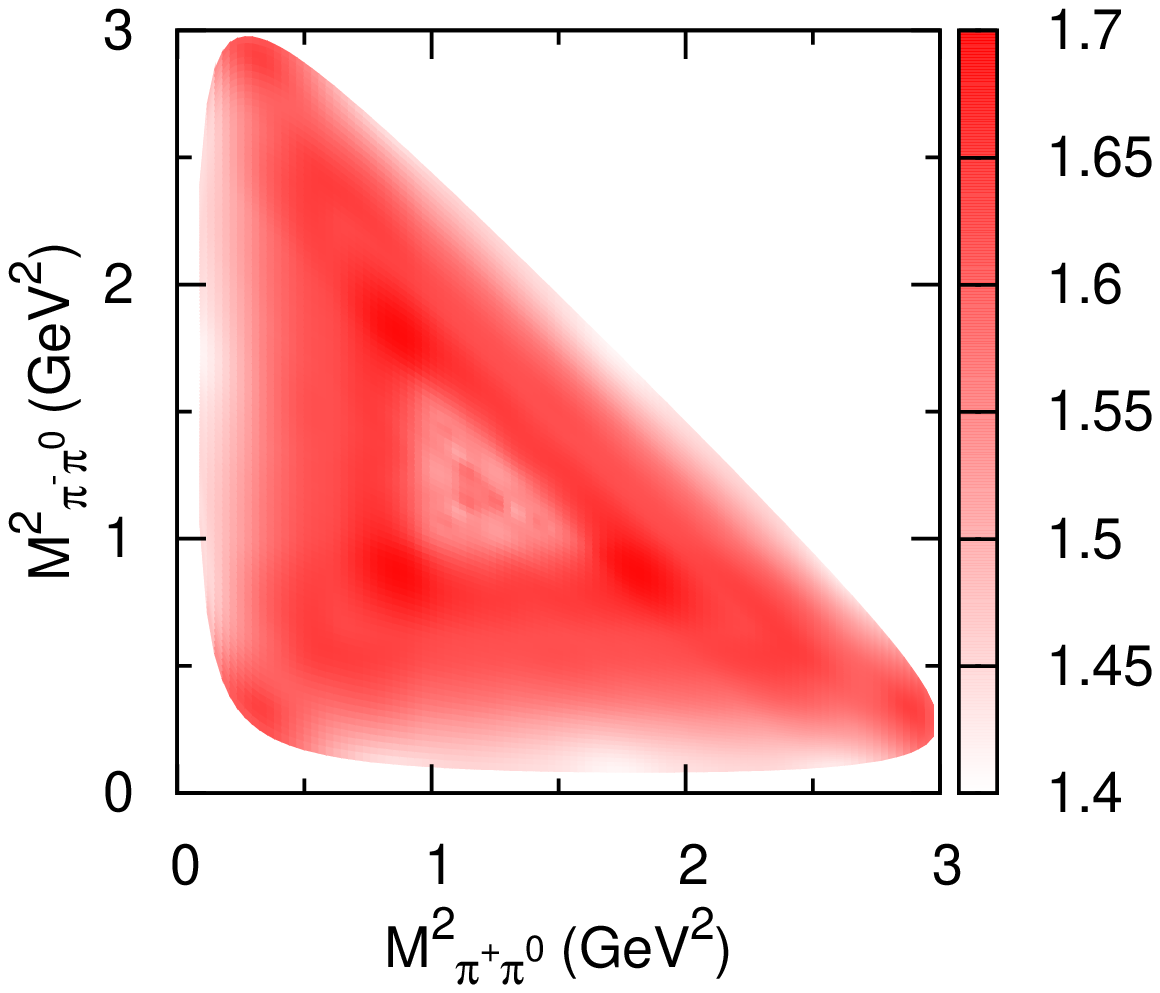}
\caption{\label{fig:D-decay}
(Color online)
(left) Dalitz plot of $D^0\to\pi^+\pi^-\pi^0$ decay; (right) Ratio of
Dalitz plot distributions with and without the $Z$ graphs.}
\end{figure}

As discussed in Sec.~\ref{sec:par-mdecay}, we only include the bare $D^0\to \pi\rho$ vertex
with  $(J^P=0^-, T=0, l=1 )$ in this calculation, as guided by
the analysis by the BABAR Collaboration~\cite{babar-d1};
the $\pi f_0$ and $\pi f_2$ channels are coupled only through the final state interaction.
The Dalitz plot calculated from our unitary amplitude 
$T_{\pi^+\pi^-\pi^0, D^0 }(E=M_{D^0}=1865~{\rm MeV})$ 
from using Eqs.~(\ref{eq:decay-tf0}) and~(\ref{eq:decay-amp-um}) is shown in Fig.~\ref{fig:D-decay} (left panel).
With an overall normalization factor, the pattern of our Dalitz plot is 
similar to BABAR's data~\cite{babar-d1}.
The sharp peaks (darker red) near the edges of distributions
are due to the formation of a $\rho$ resonance during the 3-$\pi$ propagation. 
The almost empty center part is due to the destructive interferences along 
the symmetry axes, supporting the assumption that the $T=0$ $\pi\rho$ 
channel dominates the decay~\cite{zemach}.

\begin{figure}[t]
\includegraphics[width=0.5\textwidth]{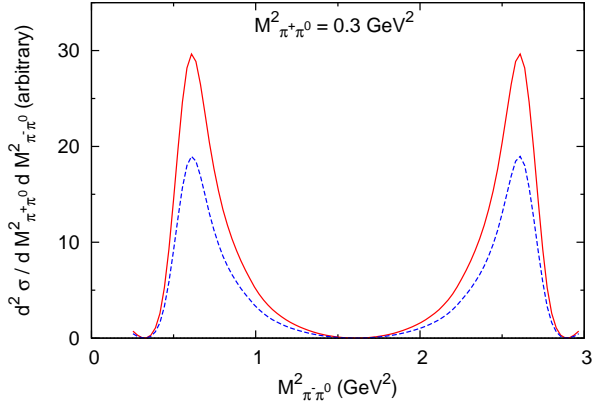}
\caption{\label{fig:D-decay-1}
(Color online)
Double differential decay width distribution [Eq.~(\ref{eq:dalitz-unpol})] of $D^0\to\pi^+\pi^-\pi^0$ decay 
at $M^2_{\pi^+\pi^0}=0.3$ (GeV)$^2$.
The red solid curve is from the full unitary model while the blue dashed curve is obtained by just 
turning off the $Z$ diagrams in the full model.
}
\end{figure}

With the same parameters and overall normalization factor,
we then calculate the Dalitz plot from $T^{\text{Isobar}}_{\pi^+\pi^-\pi^0, D^0 } ( E)$
using Eqs.~(\ref{eq:decay-tf0}) and~(\ref{eq:decay-amp-isobar}) which does not include $Z$-diagram 
mechanisms, namely, keep only the first term in Fig.~\ref{fig:mstar-decay}.
In the right panel of Fig.~\ref{fig:D-decay},
we show the ratios between the results obtained from
calculations with and without the $Z$ diagram.
Clearly, the $Z$-diagram mechanisms considerably change both 
the magnitudes and the shape of the Dalitz plot.
In most of the area, the ratios (measured by the darkness as indicated 
on the right $y$ axis of the figure) are about 1.6. 
To see this more clearly, we show in Fig.~\ref{fig:D-decay-1} 
the double differential decay width distribution, $d^2\Gamma/(dM^2_{\pi^+\pi^0} dM^2_{\pi^-\pi^0})$ defined 
in Eq.~(\ref{eq:dalitz-unpol}), at $M^2_{\pi^+\pi^0} = 0.3$ GeV$^2$. 
We see that at the $\rho$ resonance peaks,
the magnitudes can be enhanced by a factor of about 1.5 
when $Z$-diagram mechanisms are included to satisfy the three-body unitarity.

Our results shown in Figs.~\ref{fig:D-decay} and~\ref{fig:D-decay-1} indicate 
the need to reanalyze the $D$-meson decays data, with the three-body
unitarity taken into account, to
assess the results, such as CKM matrix elements, obtained with the isobar-model
analysis~\cite{babar-d0,babar-d1,babar-b6,babar-b7,belle-b6,belle-b7}.

\subsubsection{$a_1(1260), \pi_2(1670), \pi_2(2100)  \to \pi^+\pi^-\pi^0$}

The decays of these three mesons have been analyzed by using the isobar models. 
Our objective here is twofold. 
First we want to examine the $Z$-diagram effects on the Dalitz plots.
Second, we regard the Dalitz plot generated from
our unitary model [Eq.~(\ref{eq:decay-amp-um})] as the data,
and fit them with the isobar model [Eq.~(\ref{eq:decay-amp-isobar})].
We refer to it as the isobar-fit model.
In this way, we have the two models that reproduce the same Dalitz plot.
However, the decay amplitudes from the two models are not necessarily
the same, which we will examine.
This examination is particularly interesting in the context of
the extraction of the CKM phase $\gamma$ from $B$ and/or $D$ decays. 
This is because the extracted $\gamma$  depends on
the decay amplitudes, particularly on its phase\footnote{
There exists an alternative approach in which $\gamma$ can be determined
model-independently~\cite{giri,bondar1,bondar2} solely from data,
provided a large dataset is available.
A feasibility study~\cite{bondar2} showed that, with a dataset available
in the near future, the precision of $\gamma$ extracted with this approach is
comparable to that obtained with the isobar-model analysis.
Future high statistic experiments (super B factory, LHCb) make this
approach very interesting.
}.
Thus the difference in the decay amplitude 
between our unitary model and the isobar-fit model 
does matter. 
We examine this for the strong decays of $a_1$ and $\pi_2$, 
which is suggestive enough for the extraction of $\gamma$ with the isobar
model analysis.
Also, we study how well the three-body unitarity is satisfied in the
isobar-fit model.
Our unitary model satisfies it by definition (explicitly shown numerically later).
A large violation of the unitarity raises a concern about the
reliability in extracted quantities with the isobar-model analysis.

\begin{figure}[t]
\includegraphics[width=95mm]{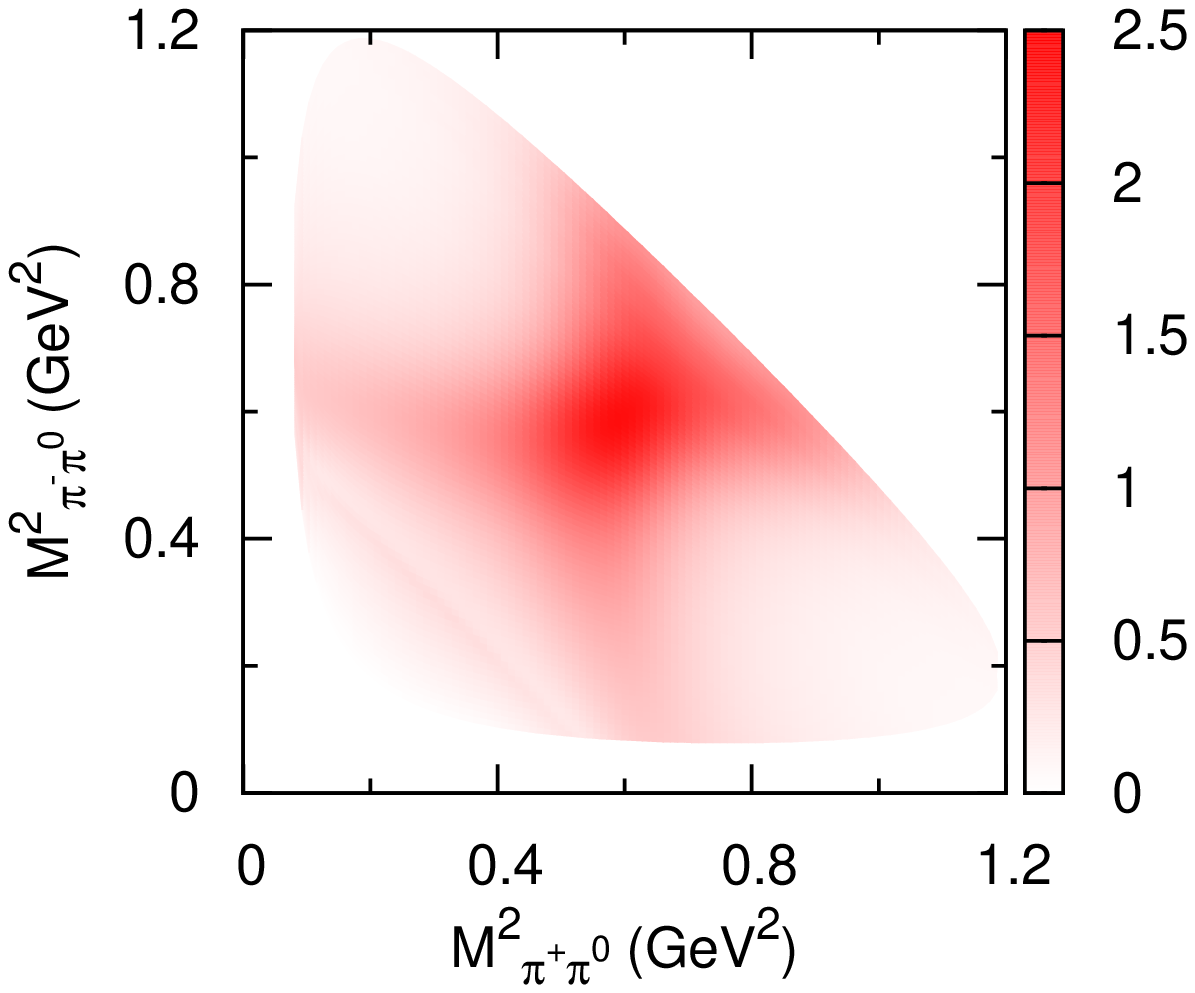}\hspace{-28mm}
\includegraphics[width=95mm]{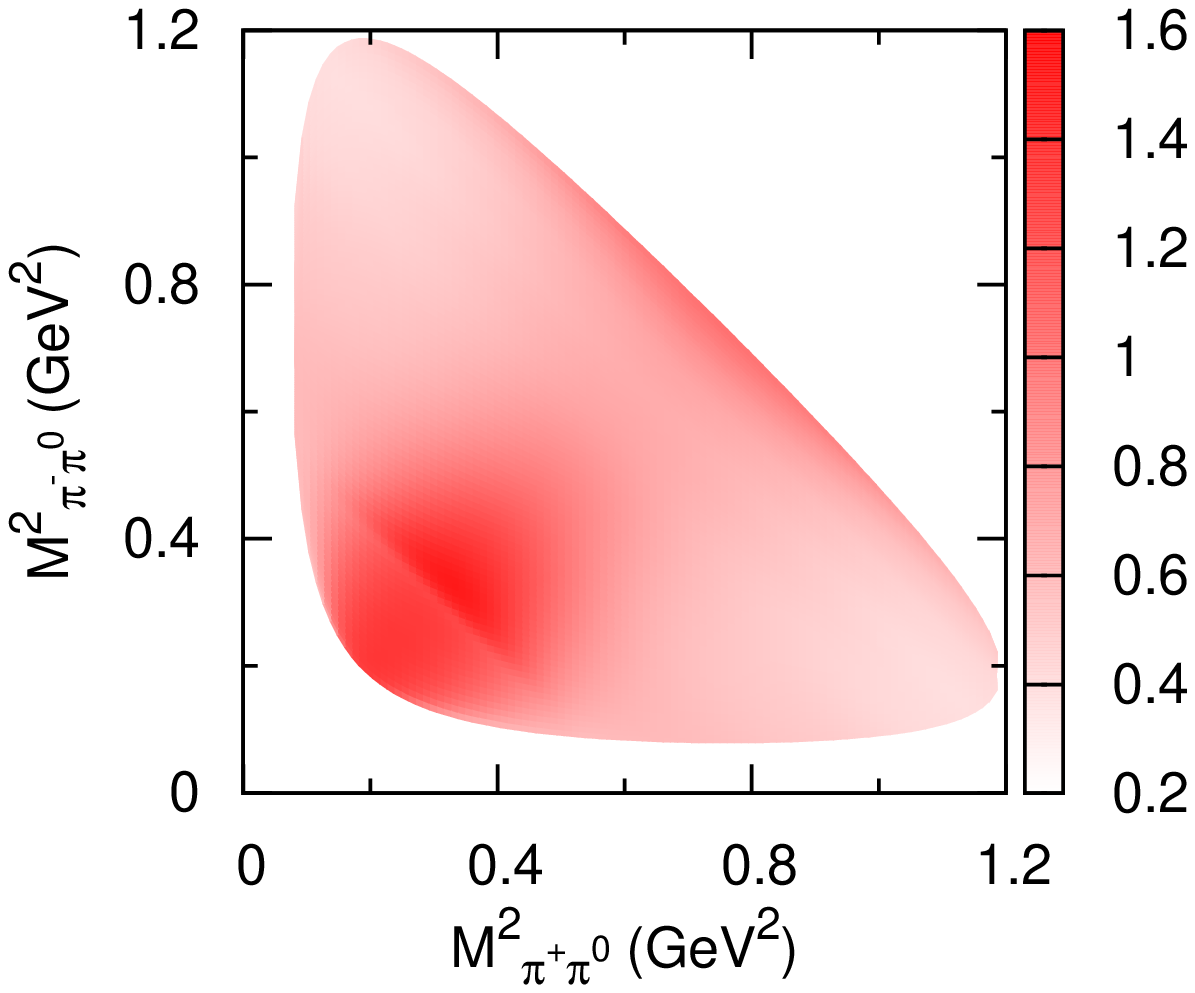}
\caption{\label{fig:a1-1260-decay}
(Color online)
(left) Dalitz plot of $a_1(1260)\to\pi^+\pi^-\pi^0$; (right) ratio of
 Dalitz plot distributions with and without the $Z$ graphs. The unit is GeV$^{-3}$.}
\end{figure}

\begin{figure}[t]
\includegraphics[width=95mm]{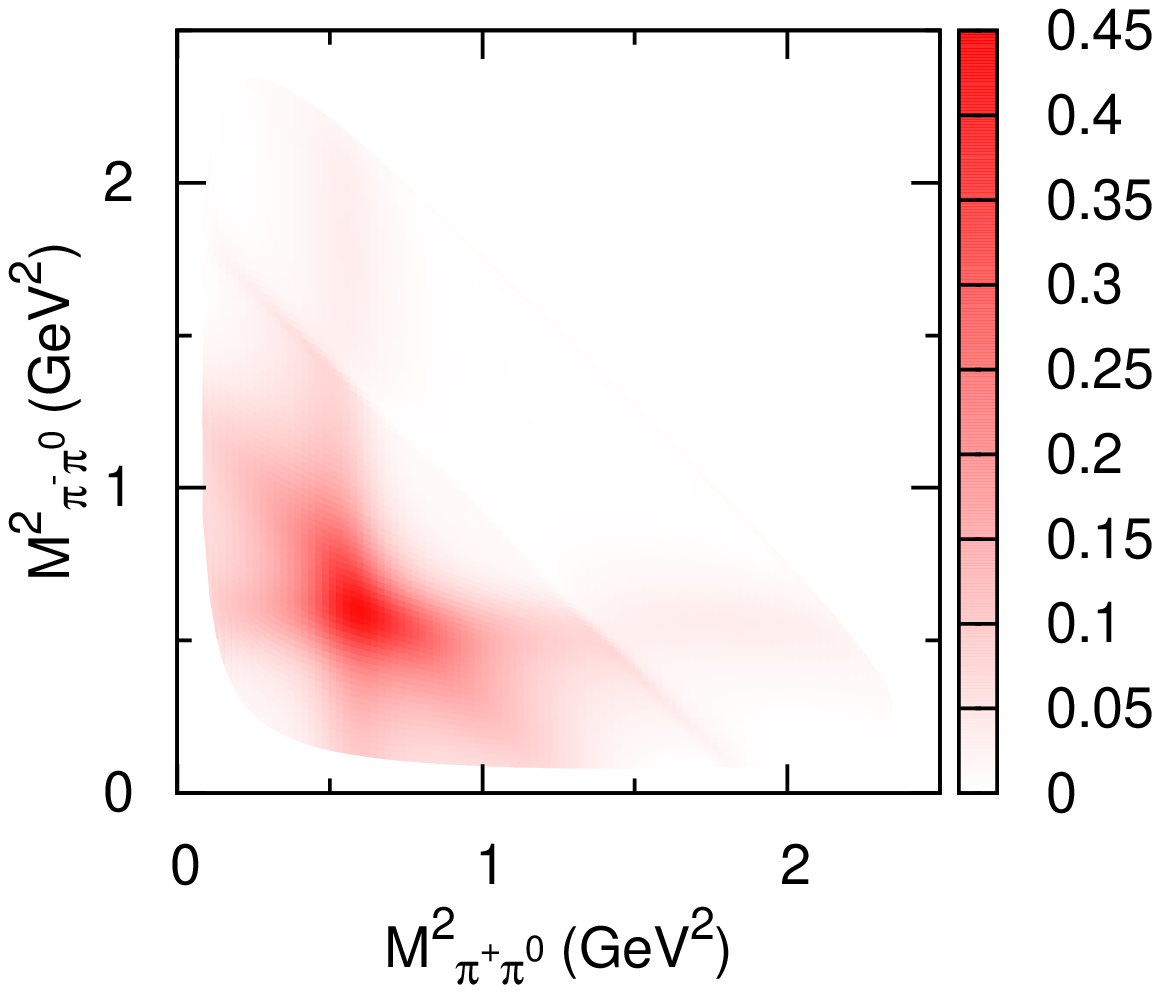}\hspace{-28mm}
\includegraphics[width=95mm]{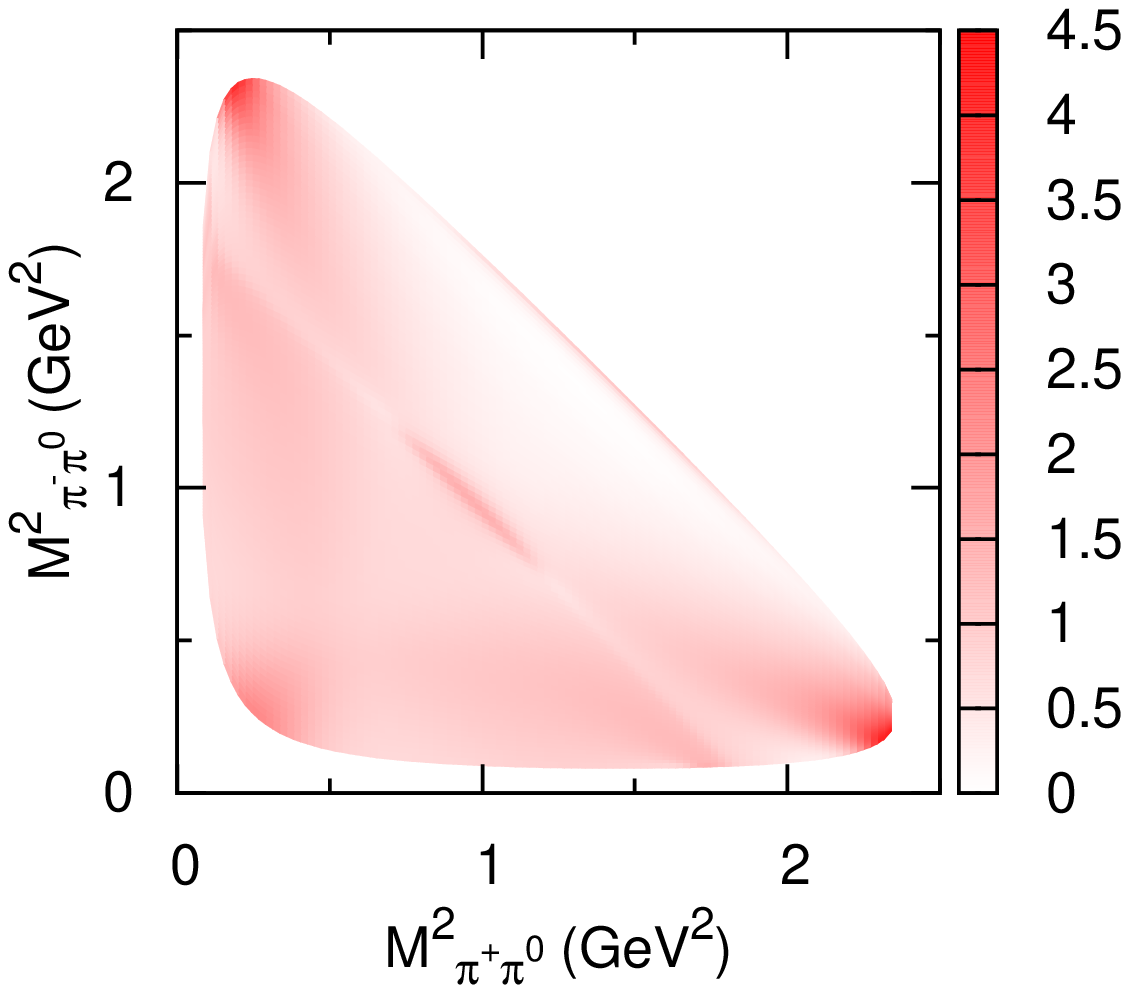}
\caption{\label{fig:pi2-1670-decay}
(Color online)
(left) Dalitz plot of $\pi_2(1670)\to\pi^+\pi^-\pi^0$; (right) ratio of
 Dalitz plot distributions with and without the $Z$ graphs. The unit is GeV$^{-3}$.}
\end{figure}

\begin{figure}[t]
\includegraphics[width=95mm]{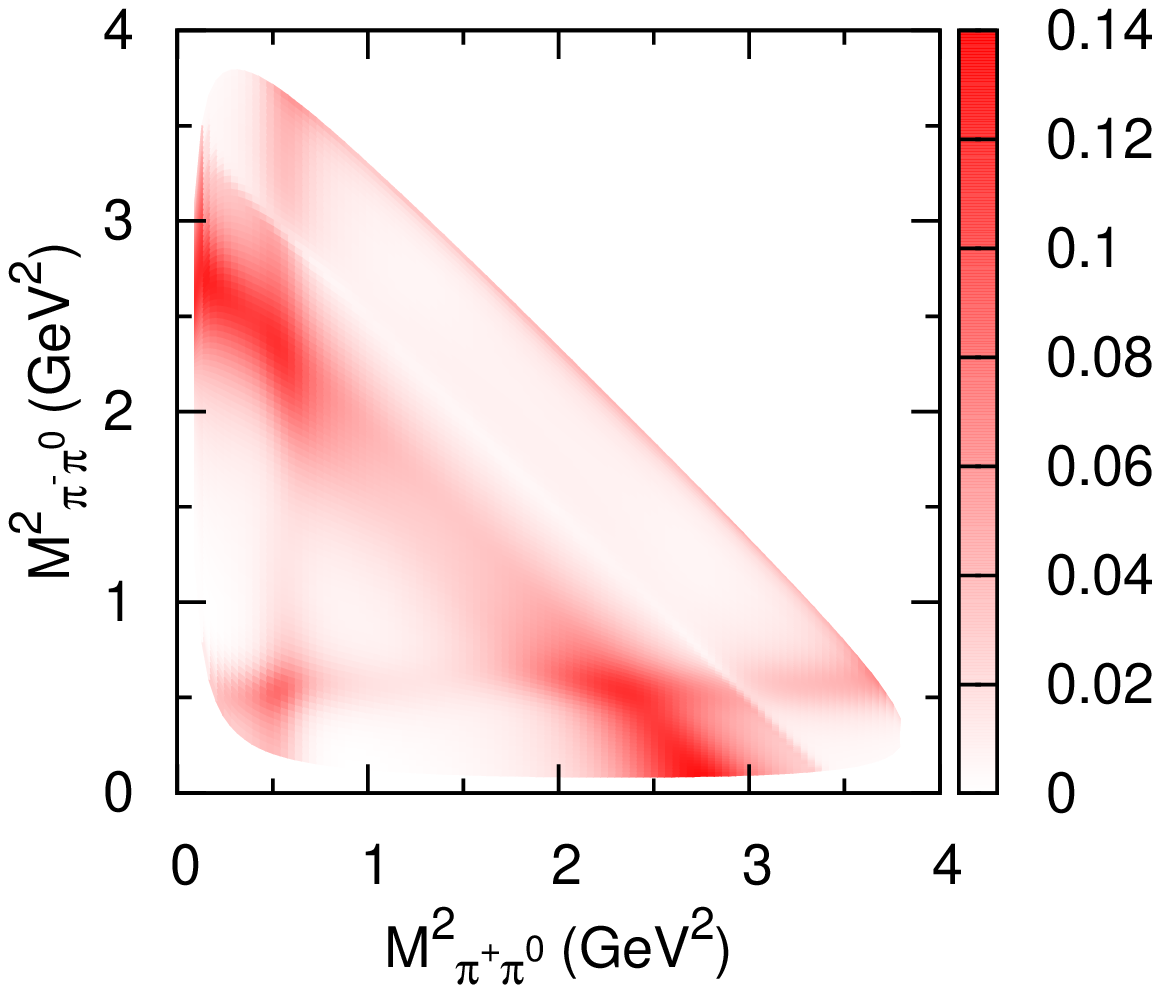}\hspace{-28mm}
\includegraphics[width=95mm]{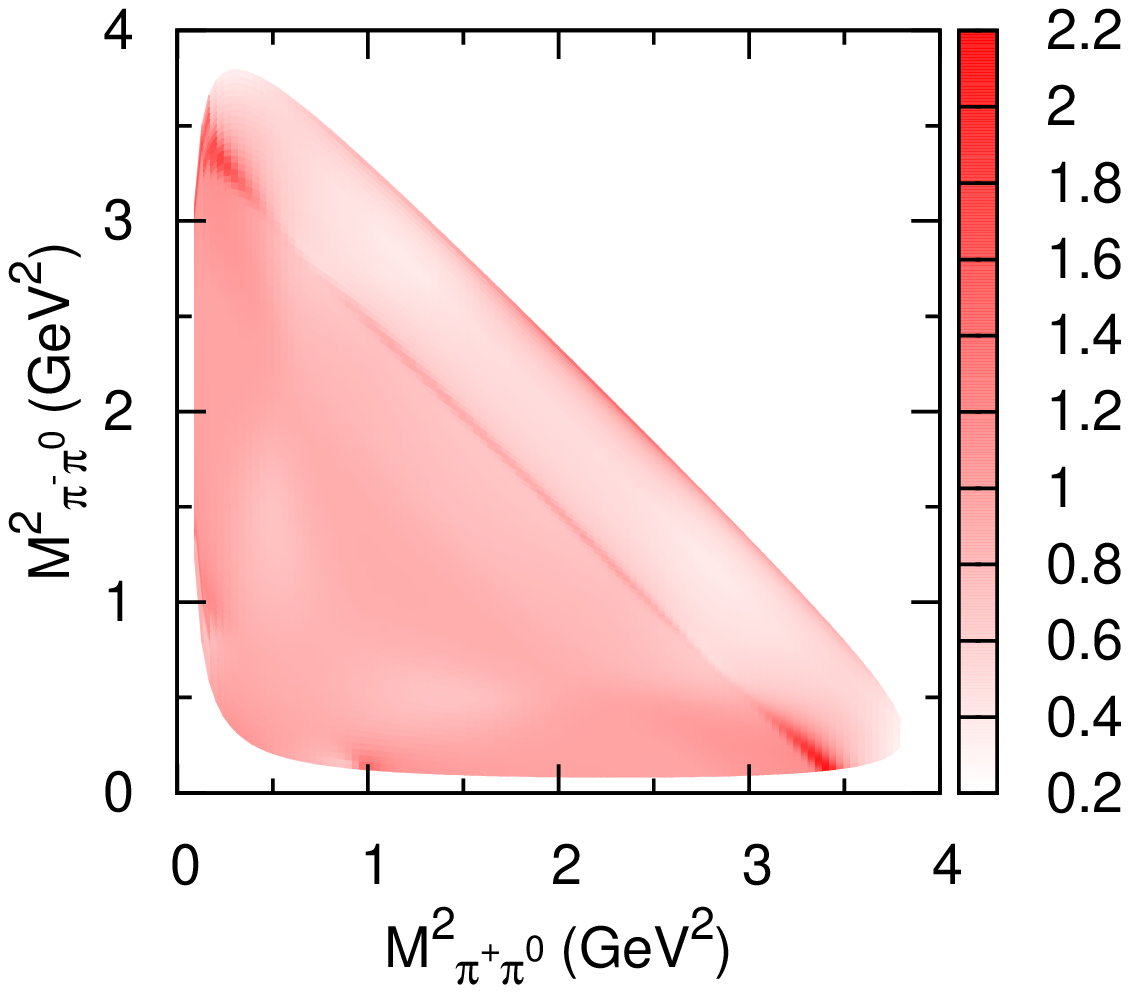}
\caption{\label{fig:dalitz}
(Color online)
(left) Dalitz plot of $\pi_2(2100)\to\pi^+\pi^-\pi^0$; (right) ratio of
 Dalitz plot distributions with and without the $Z$ graphs. The unit is GeV$^{-3}$.}
\end{figure}

We calculate the Dalitz plots at $E={\rm Re}[E_{\text{pole}}]$, for which 
we use $E_{\text{pole}}$ of the full model listed in Table~\ref{tab:pole}.
The results  from our full model [using Eq.~(\ref{eq:decay-amp-um})] are shown in
the left panels of Fig.~\ref{fig:a1-1260-decay} for $a_1(1260)$, 
Fig.~\ref{fig:pi2-1670-decay} for $\pi_2(1670)$, and Fig.~\ref{fig:dalitz} for $\pi_2(2100)$.
We see that they have rather complex structure. 
This is due to the resonances of $\pi\pi$ scattering implemented in the 
$\pi$-$R$ Green function [Eq.~(\ref{eq:green-Rc})],
and also interference among them as a consequence of summing coherently
$\pi R \to \pi\pi\pi$ partial-wave amplitudes as calculated in Eq.~(\ref{eq:decay-amp-um}).
For example, two bands on $M^2_{\pi^+\pi^0} \sim$  0.6 GeV$^2$ and
$M^2_{\pi^-\pi^0} \sim$ 0.6 GeV$^2$ in Figs.~\ref{fig:a1-1260-decay}-\ref{fig:dalitz} 
are due to the $\rho(770)$ resonance in the $p$-wave $\pi\pi$ scattering.
A gap in $M^2_{\pi^+\pi^0} + M^2_{\pi^-\pi^0}\sim$  0.6 GeV$^2$ in Fig.~\ref{fig:a1-1260-decay}
is due to the $f_0(980)$ resonance and opening of the $K\bar K$ channel.
The $Z$-diagram effects are rather different in
different parts of the Dalitz plots.
This can be seen from the ratios between the Dalitz plots calculated
with [Eq.~(\ref{eq:decay-amp-um})] and without [Eq.~(\ref{eq:decay-amp-isobar})] $Z$ diagram,  
as shown in the right-hand sides of Figs.~\ref{fig:a1-1260-decay}-\ref{fig:dalitz}.

\begin{figure}[t]
\includegraphics[width=0.5\textwidth]{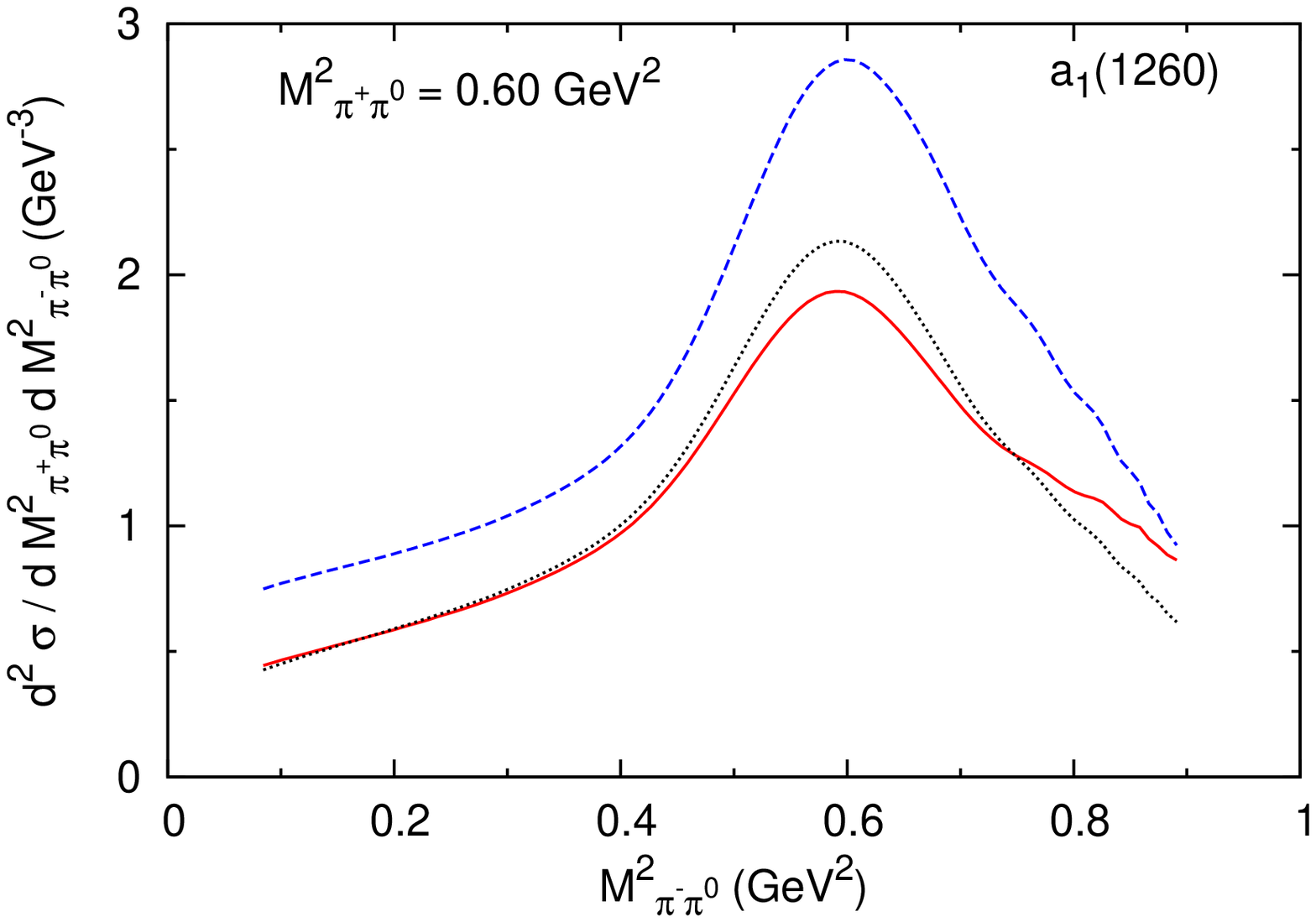}
\includegraphics[width=0.5\textwidth]{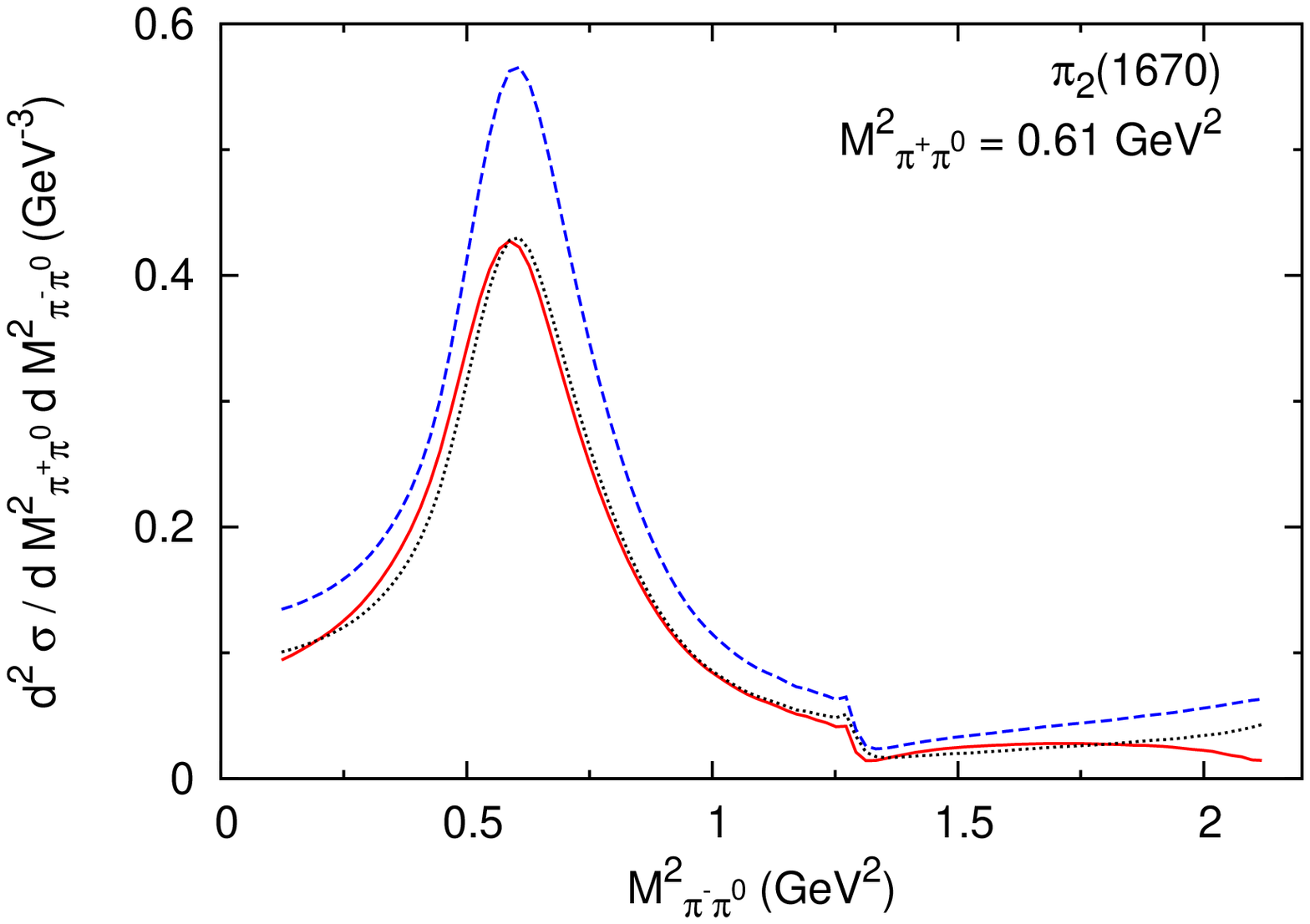}
\includegraphics[width=0.5\textwidth]{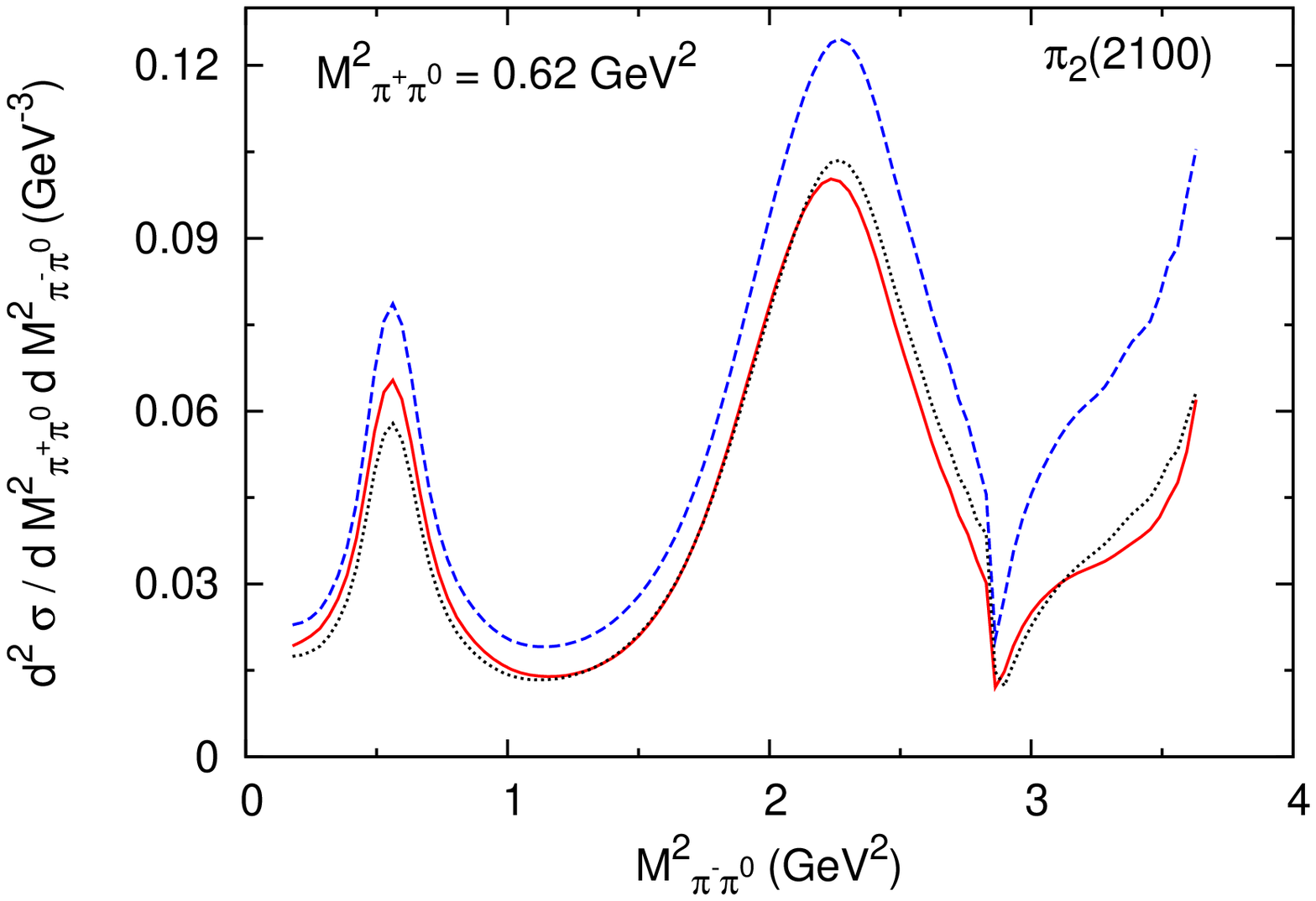}
\caption{\label{fig:a1-1260-060}
(Color online)
Double differential decay width distributions [Eq.~(\ref{eq:dalitz-unpol})]
of $a_1(1260)\to\pi^+\pi^-\pi^0$ (top),
of $\pi_2(1670)\to\pi^+\pi^-\pi^0$ (middle),
of $\pi_2(2100)\to\pi^+\pi^-\pi^0$ (bottom).
The red solid curves are from the full unitary model while the blue
dashed curves are obtained by just turning off the $Z$-diagrams in the
full model. The black dotted curves are from the isobar-fit model (see
the text for definition.).}
\end{figure}

To see the $Z$-diagram effects more clearly, we show in Fig.~\ref{fig:a1-1260-060} 
the double differential decay width distributions [Eq.~(\ref{eq:dalitz-unpol})] 
for the decays of these three mesons at typical kinematics. 
By comparing the red solid curves and blue dashed curves, we see that the
$Z$-diagram effects can significantly reduce the cross sections;
in particular in the regions near the resonance peaks of $\pi\pi$ scattering.

Now we examine differences between the unitary and
isobar models if both fit the same Dalitz plot.
We treat  the Dalitz plots in the left sides of
Fig.~\ref{fig:a1-1260-decay}-Fig.~\ref{fig:dalitz} as the data in fits using the
isobar model [Eq.~(\ref{eq:decay-amp-isobar})] by adjusting all the available
coupling constants, cutoffs of the
vertex $F_{(cR)_l,M^*}$ of Eq.~(\ref{eq:bare_mstar}). 
In the fits, we assign either $5 \%$ error for each point of the Dalitz
plot larger than 0.005~GeV$^{-3}$, or error of 0.005~GeV$^{-3}$ otherwise.
We are able to get reasonably good fits\footnote{\label{foot:isobar}
In most Dalitz plot analyses with the isobar model, a (constant)
nonresonant background amplitude with adjustable strength is included.
Also, the $M^*\to \pi R$ couplings are in general complex in the isobar model
in order to partially take
account of the missing interaction between the spectator and the paired mesons.
Because we obtained fits good enough for the following discussion, we do not
include these degrees of freedom.
}.
The resulting parameters for the isobar-fit model are rather different
from the unitary model as shown in Tables~\ref{tab:mstar-par}-\ref{tab:mstar-par-pi22100}.
The quality of our fits can be seen by comparing the dotted curves
and the solid curves in  Fig.~\ref{fig:a1-1260-060}.
Accordingly, the  decay widths to $\pi^+\pi^-\pi^0$ channel calculated from two models
using Eq.~(\ref{eq:width-cal}) 
(keeping only $abc=\pi^+\pi^-\pi^0$) agree well, as seen in 
the third, fifth and seventh columns of Table~\ref{tab:z-effect}.
However, we see in the second, fourth and sixth columns of Table~\ref{tab:z-effect} that
the resonance pole positions from resulting isobar-fit models
differ significantly from  those of the unitary model from which the Dalitz plot data are generated.
Their imaginary parts can differ by more than  100 MeV for $a_1(1260)$ and about
$50$ MeV for $\pi_2(1670)$ and $\pi_2(2100)$, indicating a large
violation of the three-body unitarity.
Note that the bare $M^*$ mass [$M^0_{M^*}$ in Eq.~(\ref{eq:mstar-g1})]
does not enter the calculation of the Dalitz plots.
Thus we choose $M^0_{M^*}$ for the isobar-fit model so that the real
part of the pole is the same as that of the unitary model.

Some hadron models predict that the hybrid mesons can have quite different branching ratios from
those of the ordinary mesons with radial excitations of the quark-antiquark pair~\cite{bcps97}.  
Thus, the parameters, $C_{\pi R, M^\ast}$, in Tables~\ref{tab:mstar-par}-\ref{tab:mstar-par-pi22100}
can provide important information to distinguish the hybrid and/or exotic mesons from 
the ordinary mesons. 
The significant difference in the parameters between the unitary and isobar models indicates 
that we should use a unitary model to analyze the Dalitz plot distributions 
to search for the exotic mesons.

\begin{table}[t]
\caption{\label{tab:z-effect}
The pole masses and total widths decaying to $\pi^-\pi^+\pi^0$ states ($\Gamma_{\pi^+\pi^-\pi^0}$) of 
$a_1(1260)$, $\pi_2(1670)$, and $\pi_2(2100)$. 
``Full'' (``Isobar fit'') is the results of the full unitary (isobar-fit) model.
}
\begin{ruledtabular}
\begin{tabular}{ccccccc}
$M^*$ & \multicolumn{2}{c}{$a_1(1260)$}& \multicolumn{2}{c}{$\pi_2(1670)$}& \multicolumn{2}{c}{$\pi_2(2100)$} \\
& Pole mass &$\Gamma_{\pi^+\pi^-\pi^0}$ & Pole mass &$\Gamma_{\pi^+\pi^-\pi^0}$ & Pole mass &$\Gamma_{\pi^+\pi^-\pi^0}$ \\
& (MeV) & (MeV) & (MeV) & (MeV) & (MeV) & (MeV) \\ \hline
Full &$1230-213i$ & 375.4 &$1672-130i$ & 157.8 &$2090-313i$ & 219.0 \\
Isobar fit &$1230-100i$ & 371.9 &$1672-~\,97i$ & 151.2 &$2090-261i$ & 217.0 
\end{tabular}
\end{ruledtabular}
\end{table}

\begin{figure}[t]
\includegraphics[width=54mm]{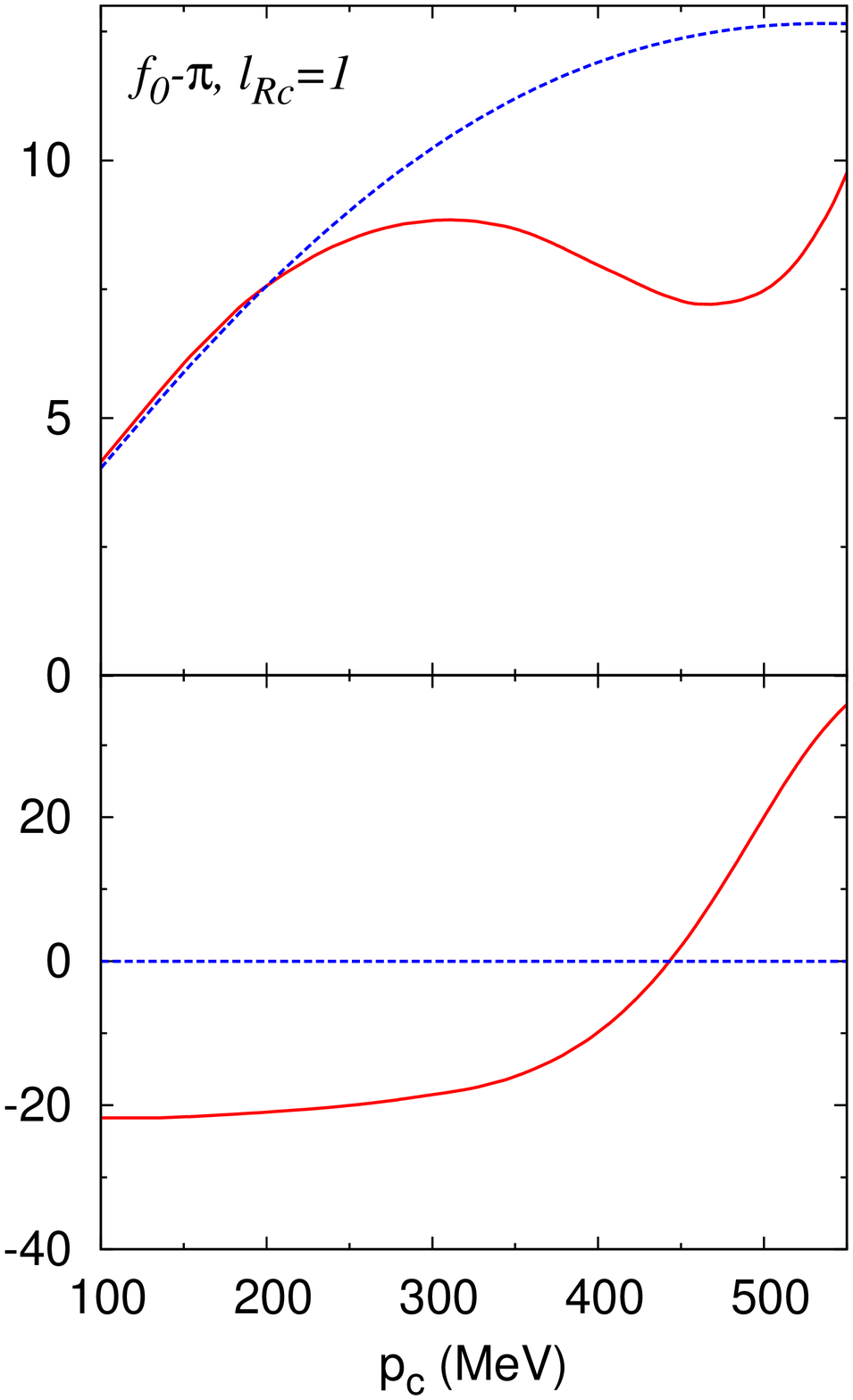}
\includegraphics[width=54mm]{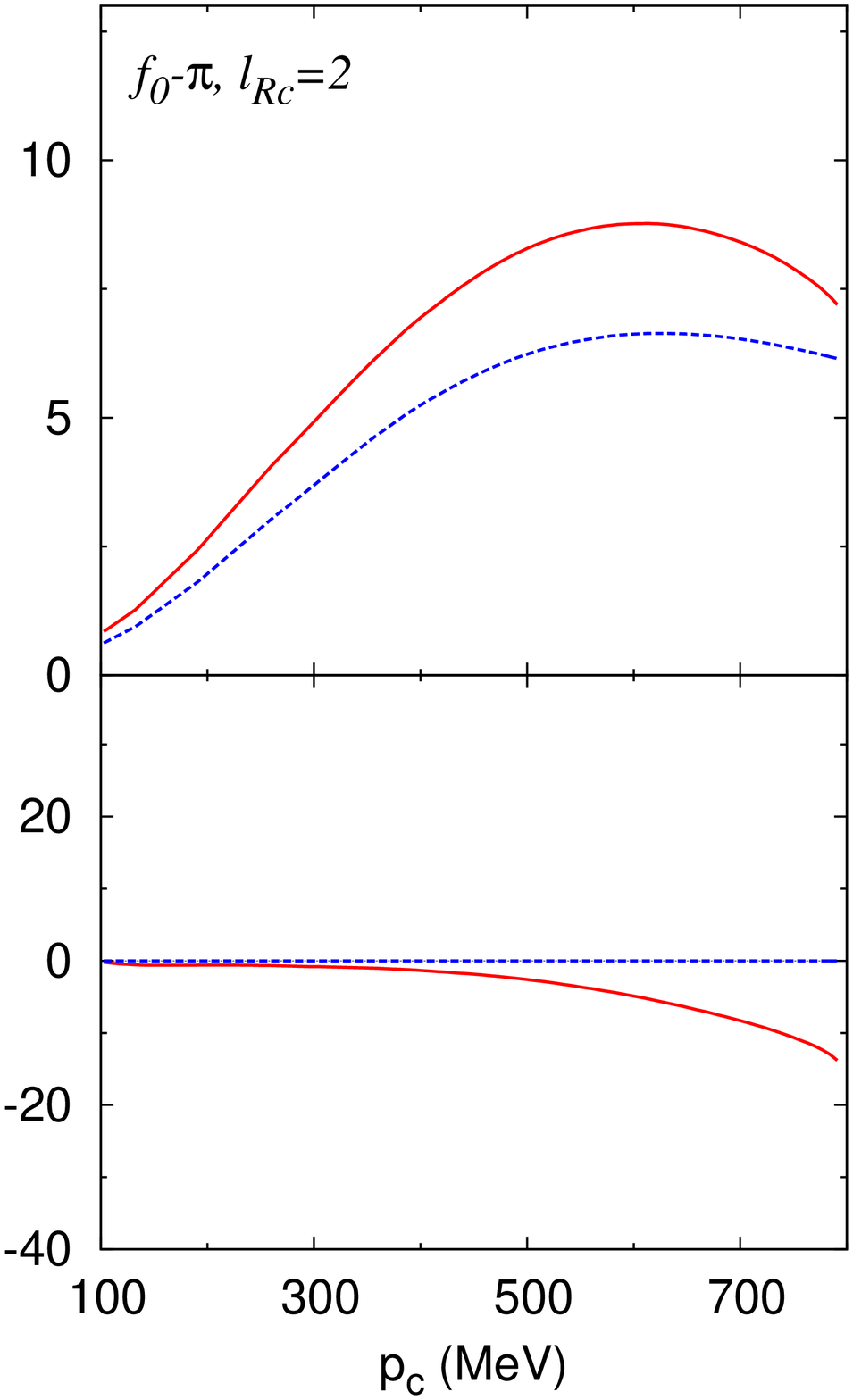}
\includegraphics[width=54mm]{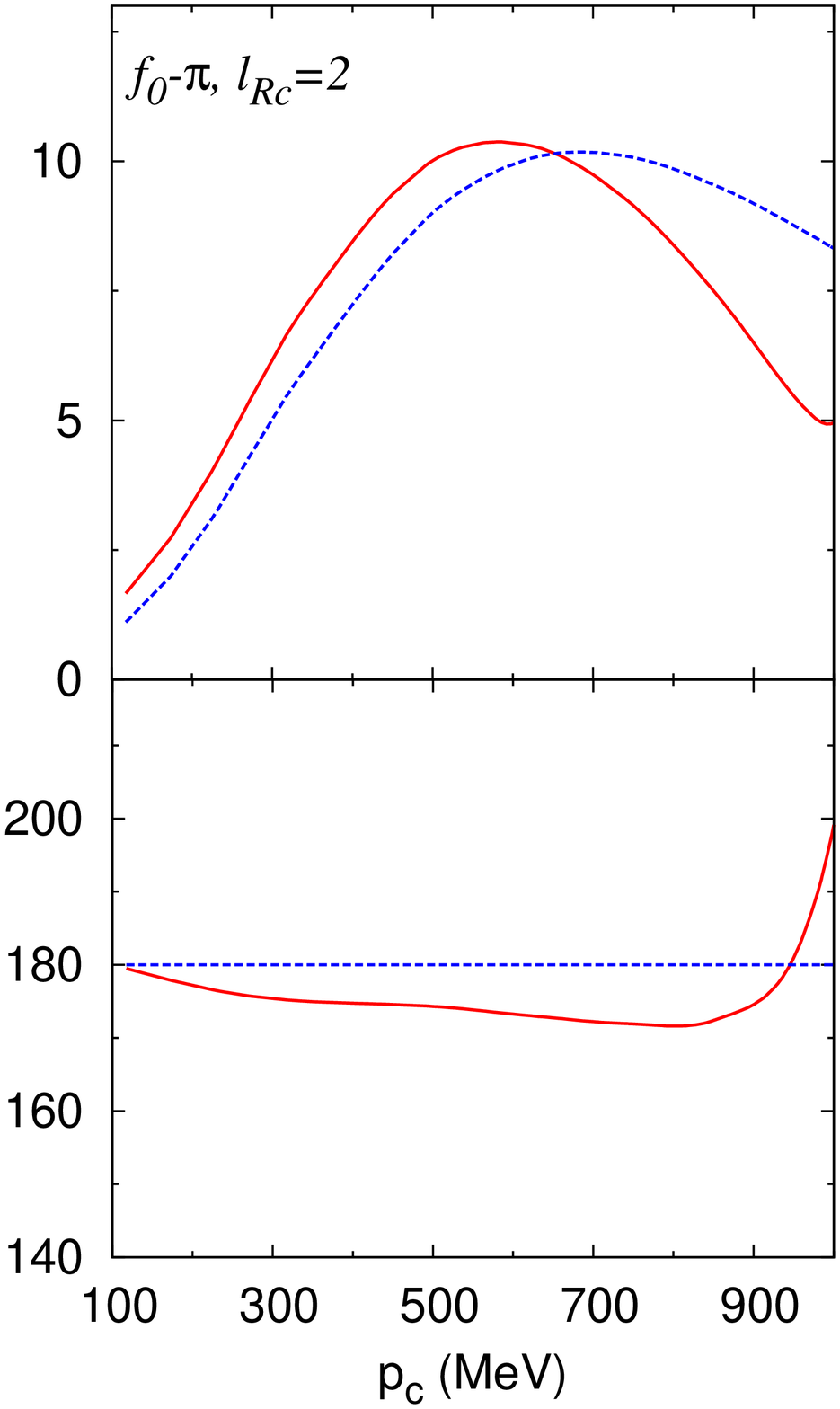}
\\ \vspace{16pt}
\includegraphics[width=54mm]{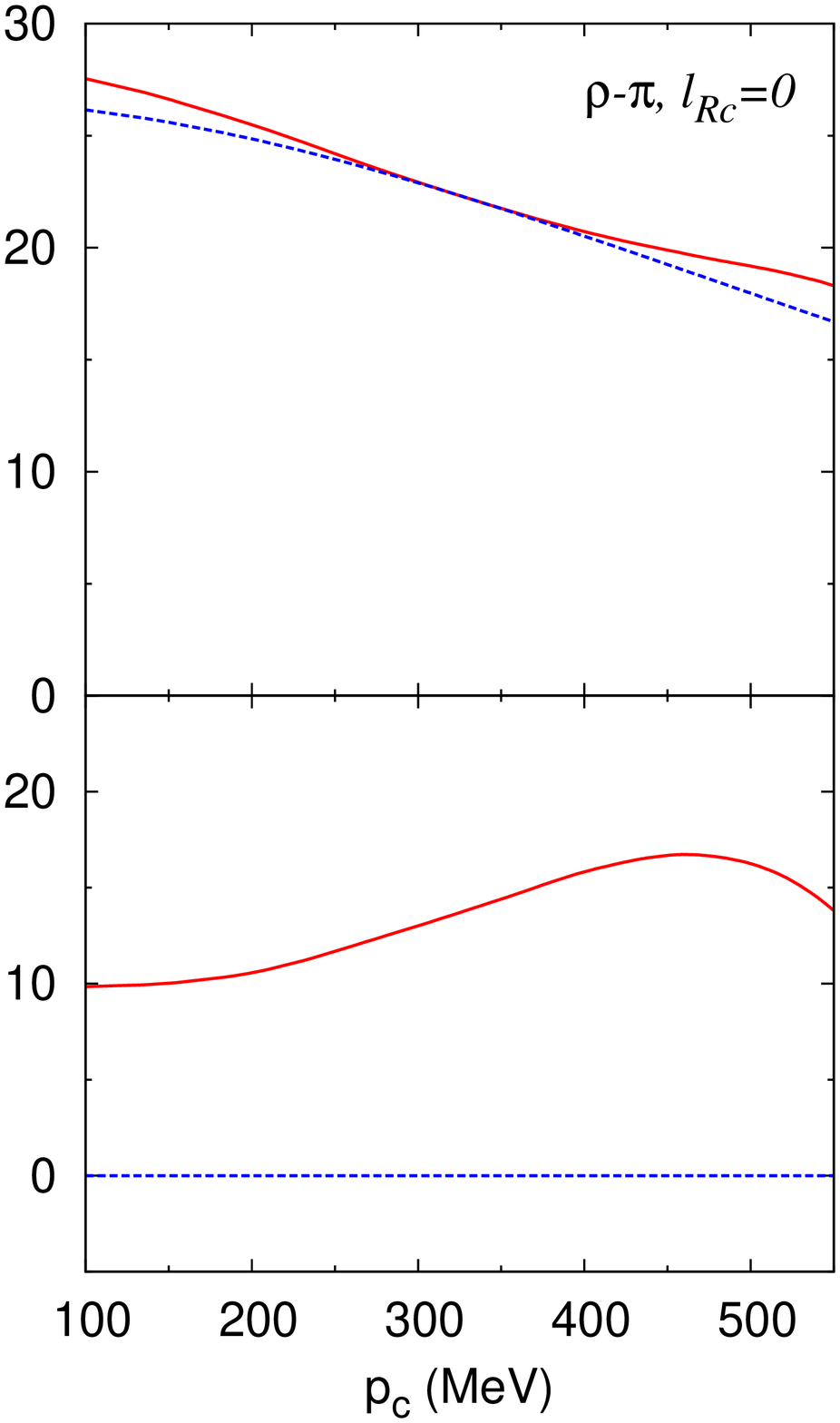}
\includegraphics[width=54mm]{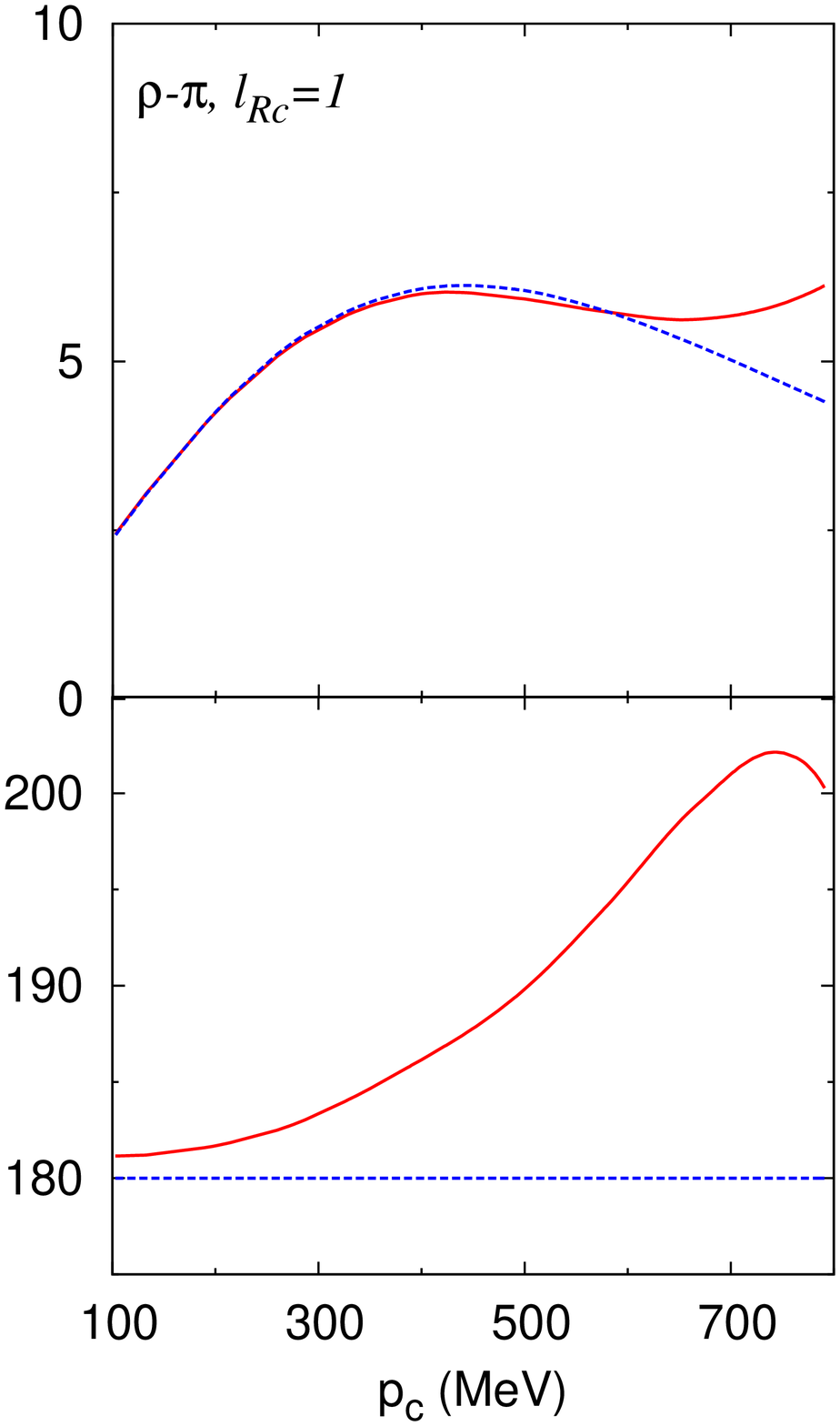}
\includegraphics[width=54mm]{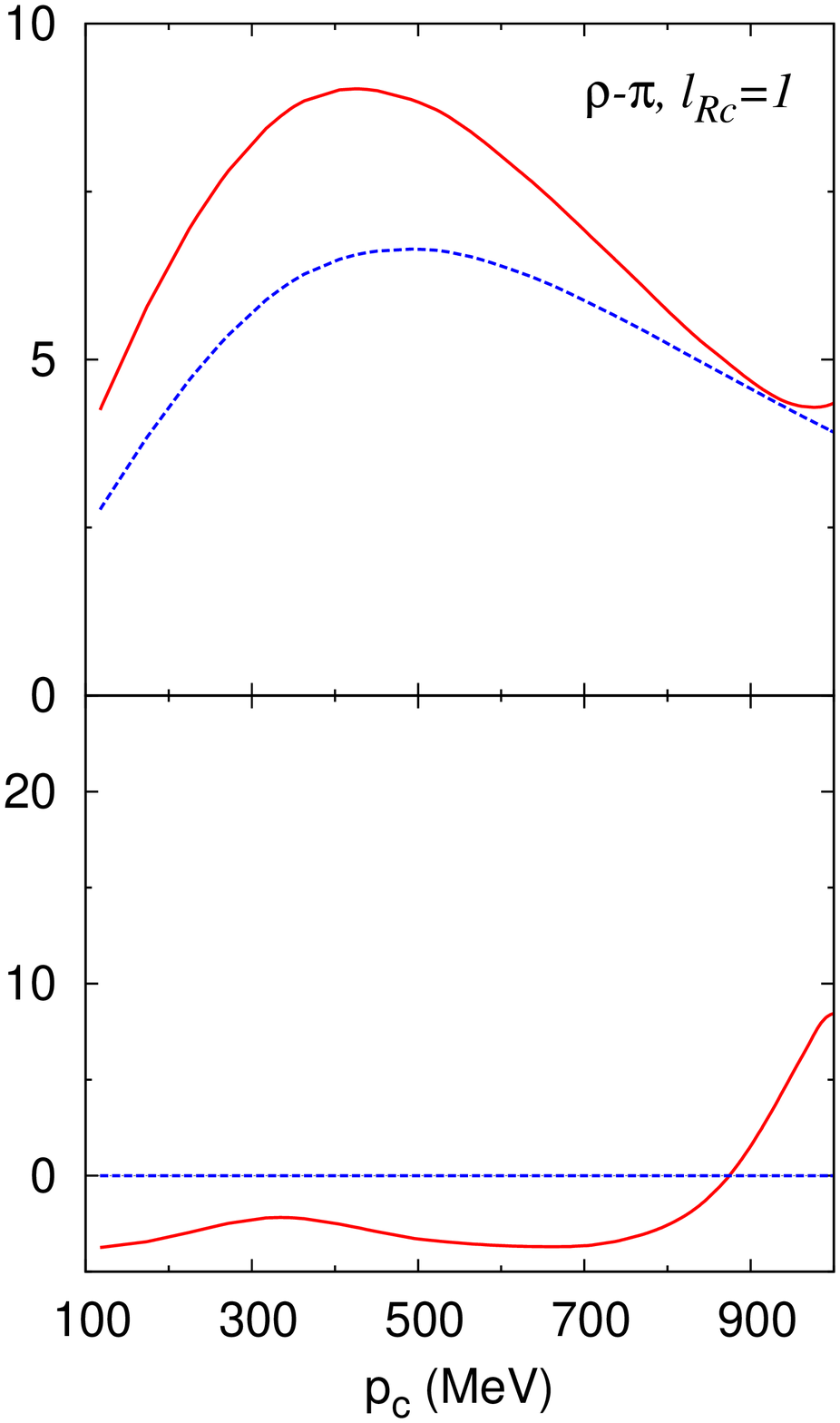}
\caption{\label{fig:mstar-vtx} 
(Color online)
The partial-wave $M^*\to cR$ vertices.
The dressed (bare) vertices of the full (isobar-fit) model
defined in Eq.~(\ref{eq:dressed-g}) [Eq.~(\ref{eq:bare_mstar})]
are shown by the red solid (blue dashed) curves. 
The upper (lower) figure shows the absolute value (phase) of the
vertices in arbitrary scale (degrees).
The figures on the left, middle and right columns are for 
$a_1(1260)$, $\pi_2(1670)$ and $\pi_2(2100)$, respectively.
The label in each figure specifies the $cR$ state.
We show the result for the bare $R=f_0$ or $\rho$ of the lowest
bare mass and $c=\pi$, and the relative orbital
angular momentum of the $cR$ is specified by $l_{Rc}$.
}
\end{figure}

As discussed above, the importance of using an unitary model can be seen more clearly in 
comparing the $M^*\to \pi R$ amplitudes predicted by the two models.
This is shown in Fig.~\ref{fig:mstar-vtx}. 
The $M^*\to \pi R$ amplitudes generated from unitary model must be complex 
because of multiple scattering due to $Z$-diagram mechanisms, while those from 
the isobar model can be chosen to be real (cf. footnote~\ref{foot:isobar}). 
Their differences in real parts can also be very different
in some regions. 
The difference in the phase is more apparent. 
Even though nonzero phase could have been used in the isobar-fit model,
as has been done in most isobar-model analyses, the rather large
dependence on the kinematics, which reflects the three-body
unitarity, is beyond the capability of the isobar model to simulate. 
As we have noted, the phases of these amplitudes are crucial
in using $D$-meson decays to determine the phase $\gamma$ of CKM matrix
elements as a way to find physics beyond the standard model.
The previously extracted $\gamma$ from
$B^{\mp} \to D^0 $ (or $\bar{D}^0$) $K^\mp \to (K_S^0 \pi^+ \pi^-) K^{\mp}$
has the uncertainty from the isobar model fitted to the $D$-decays.
It is estimated to be 8.9$^\circ$ for Belle~\cite{belle-b7}, 
3$^\circ$ for BABAR~\cite{babar-b7}.
Considering the difference in the phase, typically of 
$10^\circ\sim 20^\circ$ level, between the unitary and isobar models,
it would be highly desirable to analyze the data with the unitary model.

\begin{table}[t]
\caption{\label{tab:unitarity} Comparison between the total decay width of
bare $M^*$ ($\Gamma_{3\pi+\pi K\bar K}$)
and twice of the imaginary part of the $M^*$ self-energy ($-2 {\rm Im} [\Sigma_{M^*}]$).
Both of $\Gamma_{3\pi+\pi K\bar K}$ and $\Sigma_{M^*}$ are calculated at
$E={\rm Re}[E_{\text{pole}}]$, where $E_{\text{pole}}$ is a pole mass of a physical $M^\ast$ listed in Table~\ref{tab:z-effect}.
``Full'' (``Isobar fit'') is the results of the full unitary (isobar fit) model.
}
\begin{ruledtabular}
\begin{tabular}[b]{ccccccc}
$M^*$ & \multicolumn{2}{c}{$a_1(1260)$}& \multicolumn{2}{c}{$\pi_2(1670)$}& \multicolumn{2}{c}{$\pi_2(2100)$} \\
      &$\Gamma_{3\pi+\pi K\bar K}$  & $-2 {\rm Im} [\Sigma_{M^*}]$  &$\Gamma_{3\pi+\pi K\bar K}$  & $-2 {\rm Im} [\Sigma_{M^*}]$&$\Gamma_{3\pi+\pi K\bar K}$  & $-2 {\rm Im} [\Sigma_{M^*}]$  \\
& (MeV) & (MeV) & (MeV) & (MeV) & (MeV) & (MeV) \\ \hline
Full              & 379.8&379.7& 234.7& 234.8& 413.7& 413.4\\
Isobar fit        & 378.4&266.2& 227.4& 198.7& 434.5& 379.1
\end{tabular}
\end{ruledtabular}
\end{table}

Finally, let us examine the extent to which the three-body unitarity is
satisfied by each model. 
We can examine this using Eqs.~(\ref{eq:width-cal}) and~(\ref{eq:u-cond1})
which are satisfied by a unitary model. 
Within our current model developed for $a_1(1260)$, $\pi_2(1670)$ and $\pi_2(2100)$ decays, 
the total decay width ($\Gamma_{3\pi+\pi K\bar K}$) is the sum of 
$M^*\to 3\pi$ ($\Gamma_{3\pi}$) and $M^*\to \pi K\bar K$ ($\Gamma_{\pi K\bar K}$) widths.
In the row labeled by ``Full'' of Table~\ref{tab:unitarity}, we can 
see that the unitarity relation is satisfied within the numerical precision, as it should be.
On the other hand, for the isobar-fit model, the unitarity is rather
badly violated as seen in the fifth row of Table~\ref{tab:unitarity},
which raises a concern about the reliability of results obtained with the isobar model.

\section{\label{sec:summary}Summary and outlook}

Starting with a model Hamiltonian with vertex interactions 
$f_{ab,R}$ and $\Gamma_{cR,M^*}$ and two-body interactions $v_{c'R',cR}$, 
where $R$ and $M^*$ are the bare one-particle states and $a,b,c$ are
light pseudoscalar mesons ($\pi$, $K$, etc.), 
we have developed a unitary coupled-channels model for three-mesons decays of heavy mesons 
and excited meson states. 
By fitting the empirical amplitudes for meson-meson scattering such as $\pi\pi\to \pi\pi$,
the vertex interactions $f_{ab,R}$, 
which can generate resonances $R$ in meson-meson scattering, are determined and used 
to predict the one-particle-exchange $Z$-diagram mechanisms $Z_{c'R',cR}(E)$.
The scattering amplitudes 
$T_{c'R',cR}(E)$ are then calculated with $Z_{c'R',cR}(E)$ by solving 
a set of coupled-channels equations with the three-body unitarity condition
satisfied exactly. The  final state interactions of three-mesons from the decays 
of heavy mesons are then calculated from $T_{c'R',cR}(E)$.
In the absence of the $Z$-diagram mechanisms, our decay amplitude is reduced to
a form similar to that used in the isobar-model.
This allows us to investigate the extent to which the commonly used isobar-model 
analysis is valid in extracting the properties of heavy mesons from the Dalitz plots of 
the measured three-mesons distributions.
For strong decays of a heavy meson $M^*$, 
we present formula and procedures for investigating the importance of
three-meson interactions in determining the resonance pole positions
on complex-energy Riemann surface.

The model has been applied to investigate three-pions decays of
$a_1(1260)$, $\pi_2(1670)$ and $\pi_2(2100)$, and $D^0$ mesons. 
It was found that the $Z$-diagram mechanisms can change significantly the magnitudes and shapes of
 the  Dalitz plots. For $D^0\to \pi^+\pi^-\pi^0$, the changes in magnitudes 
can be a factor of about 1.6 in most of the phase space. 
For $a_1(1260)$, $\pi_2(1670)$ and $\pi_2(2100)$, the changes are about
a factor of $1.3 \sim 1.6$ in magnitudes 
in the regions where meson-meson resonances $f_0(600)$, $\rho(770)$ and
$f_2(1270)$ dominate.
We have also examined differences between the unitary and isobar models,
both of them produce the same Dalitz plot.
We have demonstrated that decay amplitudes from the two models are
significantly different, particularly in the phase. 
A proper estimate of the phase is particularly important for extracting
the CKM phase $\gamma$ from data of
$B^{\mp} \to D^0 $ (or $\bar{D}^0$) $K^\mp \to (K_S^0 \pi^+ \pi^-) K^{\mp}$
for which the $D$ decay Dalitz plot is analyzed with a model.
We have also shown that the three-body unitarity is rather largely violated in
the isobar model.
Finally, the resulting bare parameters which can be interpreted as characterizing 
the ``intrinsic'' quark-gluon substructure of heavy mesons are also very
different between the unitary and isobar models.

Our results strongly indicate the need for reanalysis of the $D$-meson decays using 
a unitary model to assess the results, such as the CKM matrix elements, obtained with 
the isobar-model analyses~\cite{babar-d0,babar-d1,babar-b6,babar-b7,belle-b6,belle-b7}.
It is also important to reanalyze the three-meson decays of all heavy mesons listed by PDG as  
a necessary step for establishing meson spectroscopy and exploring the hybrid or exotic mesons 
in the near future at JLab, GSI, and other possible facilities. 
While the model presented in this work is more advanced than 
the models used in the previous analyses of three-meson decays processes, 
improvements are needed to make $quantitative$ progress. 
We need to include the data associated with $K\bar{K}$ channels in determining our parameters. 
The approach for extending our formulation to include effects due to four-pions channels, 
which are considered to be important for determining  scalar-isoscalar ($L=I=0$) resonances,
should be developed. 
For $B$-meson decays, an appropriate theoretical approach must be developed 
to describe $\pi\pi$ amplitudes at high energies where no data is available.
Finally, our formulation is derived from applying a unitary transformation~\cite{kso97,sl96} to
a Hamiltonian defined within the relativistic quantum field theory. 
As discussed in Ref.~\cite{msl}, this method, as well as many well-studied three-dimensional 
reduction methods~\cite{kl74}, is needed to derive $tractable$ 
reaction models for solving complex reactions involving 
many channels and three-particle final states, with the unitarity maintained. 
Nevertheless, accuracy of these approximations should be investigated in the future.

\begin{acknowledgments}
This work is supported by the U.S. Department of Energy, Office of Nuclear Physics Division, 
under Contract No. DE-AC02-06CH11357, 
and Contract No. DE-AC05-06OR23177 under which Jefferson Science Associates operates Jefferson Lab, 
and by the Japan Society for the Promotion of Science, Grant-in-Aid for Scientific Research(C) 20540270.
H.K. acknowledges the support by the HPCI Strategic Program (Field 5
``The Origin of Matter and the Universe'') of Ministry of Education, Culture, Sports, Science
and Technology (MEXT) of Japan.
This research used resources of the National Energy Research Scientific Computing Center, 
which is supported by the Office of Science of the U.S. Department of Energy 
under Contract No. DE-AC02-05CH11231, and resources provided on ``Fusion,'' 
a 320-node computing cluster operated by the Laboratory Computing Resource Center 
at Argonne National Laboratory.
\end{acknowledgments}

\appendix

\section{\label{app:unitarity} Relation between $\Sigma_{M^*}$ and $M^*$
 decay amplitude}

Here we derive Eq.~(\ref{eq:u-cond1}).
In the course of the derivation, we will see that Eq.~(\ref{eq:u-cond1}) holds true only
when the $T$ matrix ($T'$) satisfies the unitarity relation.
We start with the ``decay'' width $\Gamma^{M^*}_{\text{tot}}(E)$ of the \textit{bare}
$M^*$ (not the physical resonance state) defined by
\begin{eqnarray}
\Gamma^{M^*}_{\rm tot}(E) = 2\pi \sum_{abc} \delta (E-E_a-E_b-E_c) |T_{abc,M^\ast} (E)|^2 .
\label{eq:gamma-2}
\end{eqnarray}
The amplitude
$T_{abc,M^\ast}$ is defined in Eq.~(\ref{eq:decay-t}), and it can be
written as
\begin{eqnarray}
T_{abc,M^\ast} = \bra{abc}(1+T'G_0) H'\ket{M^*} \ ,
\label{eq:decay-t2}
\end{eqnarray}
where $G_0$ is the free Green function, and the reaction $T$-matrix
($T'$) has been defined in Eq.~(\ref{eq:low'}).
By using Eqs.~(\ref{eq:gamma-2}) and~(\ref{eq:decay-t2}) together with the unitarity relation
\begin{eqnarray}
T' - T^{\prime\dagger} = -2 \pi i T' \delta(E-H_0) T^{\prime\dagger},
\end{eqnarray}
and the equality
\begin{eqnarray}
G_0 - G_0^\dagger = -2 \pi i \delta(E-H_0),
\end{eqnarray}
we arrive at Eq.~(\ref{eq:u-cond1}) as
\begin{eqnarray}
\Gamma^{M^*}_{\rm tot}(E) 
&=& 
2\pi \sum_{abc} \delta (E-E_a-E_b-E_c) \left|T_{abc,M^\ast} (E)\right|^2 
\nonumber\\
&=& 
2\pi \sum_{abc} \delta (E-E_a-E_b-E_c) \left|\bra{M^*}H'(1+G_0^\dagger T^{\prime \dagger})\ket{abc}\right|^2
\nonumber\\
&=& 
2\pi \bra{M^*}H'(1+G_0^\dagger T^{\prime \dagger}) \delta (E-H_0) (1+T'G_0) H'\ket{M^*}
\nonumber\\
&=& 
i \left[ \bra{M^*}H'G_0(1+T'G_0) H'\ket{M^*} 
-\bra{M^*}H'(1+G_0^\dagger T^{\prime \dagger})G_0^\dagger H'\ket{M^*} \right] 
\nonumber\\
&=& 
-2\ {\rm Im} \left[ \bra{M^*}H'G_0(1+T'G_0) H'\ket{M^*} \right] 
\nonumber \\
&=& 
-2\ {\rm Im} \left[\Sigma_{M^*}(E) \right].
\end{eqnarray}
In the last step we have used the definition of $\Sigma_{M^*}(E)$ given in Eq.~(\ref{eq:mstar-sigma}).

\section{\label{app:dalitz}Dalitz plot}

Here we summarize the formulae for calculating Dalitz plots.
The differential decay width of a heavy meson $M^\ast$ ($E$: mass; $S_{M^*}$: spin)
at rest decaying to three (pseudo) scalar mesons, 
$M^\ast(\vec{0},E)\to a (\vec{p}_a) + b (\vec{p}_b) + c (\vec{p}_c)$, where we consider
only (pseudo) scalar mesons for the final state, can be expressed as
\begin{eqnarray}
d\Gamma_{M^*}&=&
\frac{1}{2E}
\frac{d^3p_a}{(2\pi)^32E_{a}(p_a)}
\frac{d^3p_b}{(2\pi)^32E_{b}(p_b)}
\frac{d^3p_c}{(2\pi)^32E_{c}(p_c)}
\frac{{\cal B}}{2S_{M^\ast}+1}\sum_{S^z_{M^\ast}} |{\cal M}_{abc,M^\ast}|^2
\nonumber\\
&&
\times (2\pi)^4
\delta (E- E_{a}(p_a) - E_{b}(p_b) - E_{c}(p_c) )
\delta^3(\vec 0 - \vec p_a - \vec p_b - \vec p_c ),
\end{eqnarray}
where ${\cal M}_{abc,M^\ast}$ is the invariant amplitude of the decay;
${\cal B}$ is the Bose factor for the final mesons.
For example, when the three final mesons are identical, ${\cal B}=1/(3!)$.
With a variable transformation, we obtain the double differential decay width distribution
(Dalitz plot density) for the unpolarized decay given by
\begin{eqnarray}
\frac{d^2\Gamma_{M^*}}{dm_{ab}^2 dm_{bc}^2} &=&
\frac{1}{(2\pi)^3}\frac{1}{32E^3} \frac{{\cal B}}{2S_{M^\ast}+1} \sum_{S^z_{M^\ast}} |{\cal M}_{abc,M^\ast}|^2,
\label{eq:dalitz-unpol}
\end{eqnarray}
where $m_{ab}$ ($m_{bc}$) is the invariant mass of the $ab$ ($bc$) pair.
The invariant amplitude is related to the decay amplitude defined in 
Eq.~(\ref{eq:decay-tf0}) and Eq.~(\ref{eq:decay-ampf}) or Eq.~(\ref{eq:decay-amp-um}) by
\begin{eqnarray}
{\cal M}_{abc,M^\ast} = - (2\pi)^3
\sqrt{2E} \sqrt{2E_a(p_a)} \sqrt{2E_b(p_b)} \sqrt{2E_c(p_c)} T_{abc,M^\ast}.
\end{eqnarray}
The meson labels $a,b,c$
specify the momentum ($p_x$), the mass ($m_x$), and the isospin ($t_x$)
of the meson $x = a,b,c$.

Next, we summarize relations between kinematic variables.
For a given value of $m_{ab}^2$, the range of $m_{bc}^2$ is determined by its values when
$\vec p_b$ is parallel or antiparallel to $\vec p_c$:
\begin{eqnarray}
(m_{bc}^2)_{\text{max}} &=& 
(E_b^\ast +E_c^\ast)^2 - \left(\sqrt{E_b^{\ast 2} - m_b^2} - \sqrt{E_c^{\ast 2} - m_c^2}\right)^2,
\nonumber\\
(m_{bc}^2)_{\text{min}} &=& 
(E_b^\ast +E_c^\ast)^2 - \left(\sqrt{E_b^{\ast 2} - m_b^2} + \sqrt{E_c^{\ast 2} - m_c^2}\right)^2,
\end{eqnarray}
with
\begin{eqnarray}
E_b^\ast &=& \frac{1}{2m_{ab}}(m_{ab}^2-m_a^2+m_b^2),
\nonumber\\
E_c^\ast &=& \frac{1}{2m_{ab}}(E^2-m_{ab}^2+m_c^2),
\end{eqnarray}
being the energies of the particles $b$ and $c$ in the center-of-mass
frame of the ${ab}$ pair, respectively.
For a given set of $m_{ab}$ and $m_{bc}$, the momenta of the final particles are
\begin{eqnarray}
p_a &=& \frac{1}{2E}\sqrt{[E^2-(m_{bc}+m_a)^2][E^2-(m_{bc}-m_a)^2]},
\nonumber\\
p_c &=& \frac{1}{2E}\sqrt{[E^2-(m_{ab}+m_c)^2][E^2-(m_{ab}-m_c)^2]},
\nonumber\\
p_b &=& \sqrt{E_b^2-m_b^2}= \sqrt{(E-E_a-E_c)^2-m_b^2},
\nonumber\\
\cos\theta_{ab} &=& \frac{1}{2p_ap_b}\left[(E-E_a-E_b)^2-m_c^2-p_a^2-p_b^2\right],
\end{eqnarray}
where $\theta_{ab}$ is the angle between $\vec p_a$ and $\vec p_b$.
Taking $\vec{p}_a$ on the $xz$-plane, we have
\begin{eqnarray}
\vec p_a &=&p_a(\sin\theta_a,0,\cos\theta_a) ,
\nonumber\\
\vec p_b &=&
p_b(\cos\theta_a \sin\theta_{ab}\cos\phi_{ab} +\sin\theta_a \cos\theta_{ab}, \sin\theta_{ab}\sin\phi_{ab} ,
\nonumber\\
&&
\qquad
-\sin\theta_a \sin\theta_{ab}\cos\phi_{ab} +\cos\theta_a \cos\theta_{ab})  ,
\nonumber\\
\vec p_c &=&-\vec p_a -\vec p_b ,
\end{eqnarray}
where $\phi_{ab}$ is the azimuthal angle of $\vec{p}_{b'}$ which is
obtained by rotating $\vec{p}_b$ around $y$-axis by $-\theta_a$.
To calculate the differential decay width for the unpolarized decay
[Eq.~(\ref{eq:dalitz-unpol})], one may set $\theta_a=0$ and $\phi_{ab}=0$.

\section{\label{app:z}\textit{Z}-diagrams}

\subsection{Definition}

The matrix element of the $Z$-diagram for a transition process,
$R(-\vec{p}_c) + c (\vec{p}_c) \to R' (-\vec{p}_{c'}) + c'(\vec{p}_{c'})$, is given by
\begin{eqnarray}
&&
\bra{R'(-\vec{p}_{c'},s^z_{R'},t^z_{R'}) ;c'(\vec{p}_{c'},0,t^z_{c'})} Z^{c''}(E) 
\ket{R(-\vec{p}_c,s^z_{R},t^z_{R});c(\vec p_c,0,t^z_c)}
=
\nonumber\\
&&
\qquad \qquad \qquad \qquad \qquad
\bra{R'(-\vec{p}_{c'},s^z_{R'},t^z_{R'})} f_{R', c''c}
 \ket{c(\vec{p}_c,0,t^z_c); c''(\vec{p}_{c''},0,t^z_{c''})}
\nonumber\\
&&
\qquad \qquad \qquad \qquad \qquad \qquad
\times
%\left[E-E_c(p_c)-E_{c'} (p_{c'})-E_{c''}(p_{c''})+i\epsilon \right]^{-1}
\frac{1}{E-E_c(p_c)-E_{c'} (p_{c'})-E_{c''}(p_{c''})+i\epsilon} 
\nonumber \\
&&
\qquad \qquad \qquad \qquad \qquad \qquad
\times
\bra{c'\, (\vec{p}_{c'},0,t^z_{c'}); c''\, (\vec{p}_{c''},0,t^z_{c''})} 
f_{c'c'',R} \ket{R\, (-\vec{p}_c,s^z_{R},t^z_{R})}  .
\label{eq:zmx}
\end{eqnarray}
Here $c''$ is the exchanged meson; 
$s^z_{R}$ ($t^z_{R}$) is the $z$-component of the spin (isospin) of the particle $R$; 
$t^z_c$ is the $z$-component of the isospin of the particle $c$; 
$\vec{p}_{c''}=-\vec{p}_c - \vec{p}_{c'}$.
The vertices are expressed by
\begin{eqnarray}
&&
\bra{c' (\vec{p}_{c'},0,t^z_{c'}); c'' (\vec{p}_{c''},0,t^z_{c''})} f_{c'c'',R} 
\ket{R\, (-\vec{p}_c,s^z_{R},t^z_{R})} 
=
\nonumber\\
&& \qquad \qquad \qquad
J_R({p}_{c'},p_{c''},{q}_c) 
\inp{t_{c'} t^z_{c'} t_{c''} t^z_{c''}}{ t_R t^z_R}
Y_{s_R s^z_R}(\hat{q}_c)
\tilde{f}_{c'c'',R} ({q}_c) ,
\label{eq:vf-a}
\\
&&
\bra{R'\, (-\vec{p}_{c'},s^z_{R'},t^z_{R'})} f_{R', c''c}
\ket{c\, (\vec{p}_c,0,t^z_c); c''\, (\vec{p}_{c''},0,t^z_{c''})}
=
\nonumber\\
&& \qquad \qquad \qquad
J_{R'}({p}_c,p_{c''},{q}_{c'})
\inp{t_{c''} t^z_{c''}, t_{c} t^z_{c}}{ t_{R'}t^z_{R'}}
Y^*_{s_{R'} s^z_{R'}}(\hat{q}_{c'})  
\tilde{f}_{R',c''c}({q}_{c'}) .
\label{eq:vf-b}
\end{eqnarray}
The above equations are the same as Eq.~(\ref{eq:R-ab}), and $\tilde{f}$
is related to the $\pi\pi$ model 
[Eq.~(\ref{eq:pipi-vertex})] through Eq.~(\ref{eq:tilde-f}).
Here $\vec{q}_c$ is the meson momentum
in the center of mass of the two-meson ($c'c''$) subsystem from 
the $R (-\vec{p}_c)\to c' (\vec{p}_{c'}) + c'' (\vec{p}_{c''})$ decay. 
The factor $J_R$ appears as a result of the Lorentz transformation of the
vertex, and is given by
\begin{equation}
J_R(p_{x},p_{c''},{q}_y) =
\sqrt{
\frac{E_{x}({q}_y)E_{c''}({q}_y) m_R}{E_{x}(p_{x})E_{c''}(p_{c''})E_{R}(p_y)}
}  ,
\end{equation}
with $x, y=c\ {\rm or}\ c'$.
Using the Lorentz transformation, we have
\begin{eqnarray}
\vec{q}_c &=&\vec{p}_{c'} - \rho (-\vec{p}_c,\vec{p}_{c'}) \vec{p}_{c}
\equiv \kappa_c \vec{p}_c +\lambda_c \vec{p}_{c'} ,
\label{eq:qa}
\\
-\vec{q}_{c'} &=&\vec{p}_c - \rho (-\vec{p}_{c'},\vec{p}_{c}) \vec{p}_{c'}
\equiv \kappa_{c'} \vec p_{c} +\lambda_{c'} \vec p_{c'}  .
\label{eq:qb}
\end{eqnarray}
where
\begin{eqnarray}
\rho (\vec P,\vec{p}_x) =\frac{1}{\xi (\vec P,\vec{p}_x)}
\left[
\frac{\vec P\cdot \vec{p}_x}{\xi (\vec P,\vec{p}_x)
+E_{x}(p_x)+E_{c''}(|\vec P - \vec{p}_x|)} - E_{x}(p_x)
\right],
\end{eqnarray}
and
\begin{equation}
\xi (\vec P,\vec{p}_x) = 
\sqrt{\left[E_{x}(p_x)+E_{c''}(|\vec P -\vec{p}_x|)\right]^2-\vec P^2}  .
\end{equation}
The signs attached to $\vec p_{c}$ and $\vec p_{c'}$ 
in Eqs.~(\ref{eq:qa}) and~(\ref{eq:qb}), and the ordering of isospins in the 
Clebsch-Gordan coefficients in Eqs.~(\ref{eq:vf-a}) and~(\ref{eq:vf-b})
matter for the phases of the $\rho \leftrightarrow \pi\pi$ interaction,
and are taken consistently with the $\rho\pi\pi$ interaction Lagrangian.
To take these phases appropriately is important for giving the
correct phases to the amplitudes.

\subsection{Partial-wave decomposition of \textit{Z} potential}

We define the partial-wave expansion of Eq.~(\ref{eq:zmx}) as
\begin{eqnarray}
&&
\bra{R' (-\vec{p}_{c'},s^z_{R'},t^z_{R'}) ;c' (\vec{p}_{c'},0,t^z_{c'})} Z^{c''}(E) 
\ket{R (-\vec{p}_{c},s^z_{R},t^z_{R});c, (\vec p_{c},0,t^z_{c})}
=
\nonumber \\
&&
\qquad
\qquad \qquad \qquad \qquad
\sum_{TT^z}
\sum_{JJ^z}\sum_{l'l'^{z}ll^z}
\inp{t_{R} t^z_{R} t_{c} t^z_{c}}{ T T^z}
\inp{t_{R'} t^z_{R'} t_{c'} t^z_{c'}}{ T T^z}
\nonumber\\
&&
\qquad \qquad \qquad \qquad 
\qquad \qquad \qquad \qquad 
\times 
\inp{l  l^z  s_{R} s^z_{R}}{J J^z}
\inp{l' l'^z s_{R'} s^z_{R'}}{JJ^z}
\nonumber\\
&&
\qquad \qquad \qquad \qquad
\qquad \qquad \qquad \qquad
\times 
Y_{l',l'^z}(\hat{p}_{c'}) Y^*_{l,l^z}(\hat{p}_{c}) 
Z^{c'',JPT}_{(c'R')_{l'},(cR)_l}(p_{c'},p_{c}) .
\label{eq:pw-expansion}
\end{eqnarray}
Performing some manipulations, we obtain
\begin{eqnarray}
Z^{c'',JPT}_{(c'R')_{l'},(cR)_l}(p_{c'},p_{c}) &=&
(-1)^{t_{c''}-t_{R}+t_{c'}}
\sqrt{(2t_{R}+1)(2t_{R'}+1)} 
W(t_{c} t_{R'} t_{R} t_{c'};t_{c''} T)
\nonumber\\
&&\times
(-1)^{s_{R'}}
\sqrt{(2s_{R'}+1)(2s_{R}+1)(2l'+1)(2l+1)}
\nonumber\\
&&\times
\sum_{l_a,l_b,L',L'',j}
(2j+1)(2L'+1)(2L''+1)
\sqrt{\frac{(2s_{R'}+1)!(2s_{R}+1)!}{(2l_a)!(2s_{R'}-2l_a)!(2l_b)!(2s_{R}-2l_b)!}}
\nonumber\\
&&\times
\left(
\begin{array}{ccc}
l'&l_a&L'\\
0  &0 &0
\end{array}
\right)
\left(
\begin{array}{ccc}
l&l_b&L''\\
0  &0 &0
\end{array}
\right)
\left(
\begin{array}{ccc}
s_{R}-l_b&j&L'\\
0  &0 &0
\end{array}
\right)
\left(
\begin{array}{ccc}
s_{R'}-l_a&j&L''\\
0  &0 &0
\end{array}
\right)
\nonumber\\
&&\times
\left\{
\begin{array}{ccc}
l'&l_a&L'\\
s_{R'}-l_a&J&s_{R'}\\
\end{array}
\right\}
\left\{
\begin{array}{ccc}
l&l_b&L''\\
s_{R}-l_b&J&s_{R}\\
\end{array}
\right\}
\left\{
\begin{array}{ccc}
L''&s_{R}-l_b&J\\
L'&s_{R'}-l_a&j\\
\end{array}
\right\}
F^{l_a,l_b}_j  ,
\label{eq:zj-final}
\end{eqnarray}
where we have used $(-1)^{l+l'+s_{R}+s_{R'}}=1$
from the parity conservation.
We have introduced $F^{l_a,l_b}_j$ and $B$ defined by
\begin{eqnarray}
F^{l_a,l_b}_j &=& 
\frac{1}{2} \int^1_{-1} dx
\frac{BP_j(x)}{E-E_{c}(p_{c})-E_{c'} (p_{c'})-E_{c''}(p_{c''})+i\epsilon}  ,
\end{eqnarray}
\begin{eqnarray}
B
&=&
J({p}_{c},p_{c''},{q}_{c'}) \tilde{f}^{s_{R'}t_{R'}}_{cc'', R'}({q}_{c'}) 
J({p}_{c'},p_{c''},{q}_{c}) \tilde{f}^{s_Rt_R}_{R, c' c''}({q}_{c}) 
\nonumber \\
&&\times
(\lambda_{c'} p_{c'})^{l_a} (\kappa_{c'} p_{c})^{s_{R'}-l_a} 
(\lambda_{c} p_{c'})^{s_{R}-l_b} (\kappa_{c} p_{c})^{l_b} 
(q_{c})^{-s_{R}} (q_{c'})^{-s_{R'}} ,
\end{eqnarray}
where $P_j(x)$ is the Legendre function of the degree $j$.

%\bibliography{3pi-paper}


\end{document}